# Logic Bug Detection and Localization Using Symbolic Quick Error Detection

Eshan Singh, David Lin, Clark Barrett, and Subhasish Mitra, *Fellow, IEEE*

***Abstract*—We present Symbolic Quick Error Detection (Symbolic QED), a structured approach for logic bug detection and localization which can be used both during pre-silicon design verification as well as post-silicon validation and debug. This new methodology leverages prior work on Quick Error Detection (QED) which has been demonstrated to drastically reduce the latency, in terms of the number of clock cycles, of error detection following the activation of a logic (or electrical) bug. QED works through software transformations, including redundant execution and control flow checking, of the applied tests. Symbolic QED combines these error-detecting QED transformations with bounded model checking-based formal analysis to generate minimal-length bug activation traces that detect and localize any logic bugs in the design.**

**We demonstrate the practicality and effectiveness of Symbolic QED using the OpenSPARC T2, a 500-million-transistor open-source multicore System-on-Chip (SoC) design, and using "difficult" *logic bug* scenarios observed in various state-of-the-art commercial multicore SoCs. Our results show that Symbolic QED: (i) is fully automatic, unlike manual techniques in use today that can be extremely time-consuming and expensive; (ii) requires only a few hours in contrast to manual approaches that might take days (or even months) or formal techniques that often take days or fail completely for large designs; and (iii) generates counter-examples (for activating and detecting logic bugs) that are up to 6 orders of magnitude shorter than those produced by traditional techniques. Significantly, this new approach does not require any additional hardware.**

## I. Introduction

With the rapidly growing complexity of integrated circuits (ICs), critical design flaws (bugs) frequently escape pre-silicon verification, when the implemented design is checked to see if it meets defined specifications prior to the system being physically fabricated. As a result, there is an increasing dependence on post-silicon validation of manufactured ICs in actual system environments to detect and fix these bugs. Design bugs can be broadly classified into:

1. *Logic bugs* that are caused by (logic) design errors (including incorrect interactions between the hardware implementation and low-level system software, e.g. firmware);
2. *Electrical bugs* that are caused by subtle, often non-deterministic, interactions between a design and its "electrical" state. **This paper focuses on logic bugs.**

Existing *pre-silicon* verification is inadequate for "difficult" logic bugs. Critical design bugs escape pre-silicon verification and are detected only during post-silicon validation [1-6]. Existing post-silicon validation and bug localization practices are often manual and generally *ad hoc*, and their costs are rising faster than design costs [2, 7-9]. These "difficult" bugs are, by definition, challenging to activate and detect during pre-silicon verification. In manufacturing test, well-established test coverage metrics (e.g. single-stuck-at coverage, transition fault coverage) exist. These metrics have been experimentally shown to be effective in detecting manufacturing defects. Metrics such as code coverage and assertion coverage are used during pre-silicon verification, but are less standardized. For post-silicon validation, coverage metrics are in their infancy and are highly challenging (partially due to very limited controllability and observability with "traditional" post-silicon validation).

Following error detection, *post-silicon bug localization* involves identifying a *bug trace* (defined as a sequence of inputs, e.g., instructions that activate and detect the bug) and the hardware design block (possibly) containing the bug. The effort to localize bugs from observed system failures (e.g., deadlocks, crashes, output errors) dominates the overall cost of post-silicon validation and debug [2, 4, 8, 10]. For example, it might take days or weeks (or even months) of (manual) work to localize and debug a single logic bug [4, 11]. New techniques are needed to reverse this trend.

Post-silicon bug localization challenges are primarily caused by long error detection latencies [12-15]. *Error detection latency* is the time elapsed between when a test activates a bug and creates an error and when the error manifests as an observable failure (e.g., system crash, timeout, deadlock, exception). Error detection latencies for "difficult" bugs can exceed several millions or even billions of clock cycles where the error remains latent and does not affect an observed signal [13, 14]. It is extremely difficult to trace that far back into the history of system operation, especially for large designs consisting of multiple cores, cache / memory controllers, etc.

Traditional post-silicon validation and debug techniques often rely on trace buffers to generate bug traces. *Trace buffers* are small memories that record the logic values of a selected set of signals. Typically, trace buffers can record only a few (~1,000) clock cycles of history (or a longer history at the cost of recording fewer signals) [7, 16, 17]. However, when dealing with extremely long error detection latencies (especially for multi-core chips with many signals to record), trace buffer techniques can quickly become ineffective.

Assertions can also be used for post-silicon debug. However, manual assertion creation is difficult, and it is even more challenging to create assertions that can be efficiently implemented in hardware. Although reconfigurable logic has been shown to reduce the implementation burden [7], selecting the "right" set of assertions to include remains a major problem. This issue is a significant challenge facing automatic assertion generation [18-21], which can see an explosion in the number of assertions, many of which are ineffective at catching bugs.

Many existing bug localization practices rely on failure reproduction, which involves returning the system to an error-free state and re-executing the failure-causing stimuli. As explained in [16, 22], failure reproduction is very difficult for complex ICs due to non-deterministic behaviors, such as interrupts, I/O functionalities, interactions between multiple processor cores, and operating system functionalities (e.g., context switches). The sheer design size also poses major challenges. System-level simulations are several orders of magnitude slower than actual silicon [1, 4, 23]. The use of formal analysis and Boolean Satisfiability techniques for post-silicon validation and debug (e.g., [16, 24, 25]) can also be severely limited by design size (as we also show in Sec. IV).

*Symbolic Quick Error Detection* (*Symbolic QED*) is motivated by the urgent need for a structured, automated, and scalable approach to overcome post-silicon bug localization challenges. However, since it only requires the design RTL, it is also directly applicable for detecting and localizing logic bugs during pre-silicon verification (we introduce an important post-silicon application in Sec. IV.D to detect bugs that escape pre-silicon). Key characteristics of Symbolic QED are: 1) It is applicable to any System-on-Chip (SoC) design

E. Singh and D. Lin are with the Department of Electrical Engineering, Stanford University, Stanford, CA 94305 USA.

C. Barrett is with the Department of Computer Science, Stanford University, Stanford, CA 94305 USA.

S. Mitra is with the Department of Electrical Engineering and Department of Computer Science, Stanford University, Stanford, CA 94305 USA.

containing at least one programmable processor core (a generally valid assumption for existing SoCs [3]); 2) It is broadly applicable for logic bugs inside processor cores, accelerators, and uncore components;[1] 3) It doesn't require failure reproduction; 4) It doesn't require human intervention during bug localization; 5) It doesn't require additional hardware to localize logic bugs. However, small hardware blocks called change detectors can improve localization during post-silicon validation and debug; and, 6) It doesn't require design-specific assertions.

We demonstrate the effectiveness and practicality of Symbolic QED by showing that: 1) Symbolic QED correctly and automatically localizes difficult logic bugs in a few hours (less than 7) for OpenSPARC T2, a 500-million-transistor open-source SoC (see Sec. IV). Such bugs would generally take days or weeks (or even months) of manual work to localize using traditional approaches; 2) Symbolic QED does not require additional hardware (such as trace buffers) for localizing logic bugs; 3) For each detected logic bug, Symbolic QED provides a small list of candidate components representing the possible locations of the bug in the design; 4) For each detected logic bug, Symbolic QED automatically generates a minimal-length bug trace using formal analysis; 5) Bug traces generated by Symbolic QED are up to 6 orders of magnitude shorter than those produced by traditional techniques; and, 6) while Symbolic QED is motivated in the context of post-silicon validation and debug, it does not require any actual tests run on silicon, making it applicable to pre-silicon design verification.

Our first paper [26] introduced the Symbolic QED methodology using the EDDI-V and PLC QED transforms, and demonstrated its effectiveness using "difficult" bug scenarios abstracted from the bug databases of commercial multi-core SoCs. This paper builds on that work in the following ways:

1. We extend Symbolic QED to more QED transformations: Control Flow Checking using Software Signatures for Validation (CFCSS-V), and Control Flow Tracking using Software Signatures for Validation (CFTSS-V).
2. We expand on the results from [26] to include control flow bugs, including an example causing a deadlock.
3. We provide more detailed analysis and discussions on how Symbolic QED handles large designs, difficult bugs that feature system interrupts, and bugs that would otherwise require locks but do not with our approach.
4. We demonstrate how Symbolic QED can be used during post-silicon validation to detect and localize bugs (undetected in pre-silicon) with very long activation sequences that require a specific BMC starting state, when combined with QED tests.

### A. *Motivating Example*

We present a bug scenario that corresponds to a difficult bug found during post-silicon validation of a commercial multicore SoC:

*Two stores within 2 cycles to adjacent cache lines delay the next cache coherence message received by that cache by 5 clock cycles.*

The bug is *only* activated when two store operations to adjacent cache lines occur within 2 clock cycles of each other. The next cache coherence message (e.g., invalidation) is delayed because of a delay in the receive buffer of the cache (these details were not known before the bug was found and localized). During post-silicon validation, a test running on the SoC created a deadlock. As shown in Fig. 1, the deadlock occurred because one of the processor cores (core 4) performed a store to memory location [A] followed by a store to memory location [B] within 2 cycles ([A] and [B] were cached on adjacent cache lines). As a result, the bug was activated in cache 4.

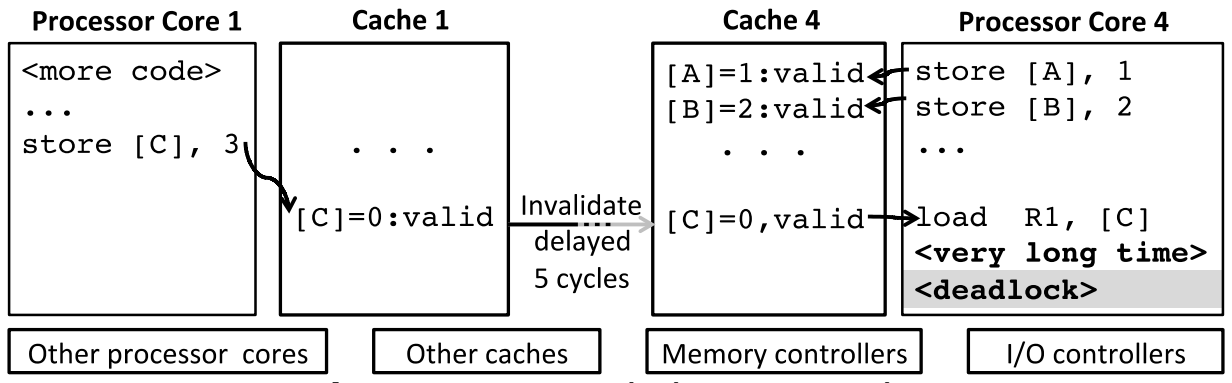

**Figure 1.** Example bug scenario.

After the bug was activated, processor core 1 performed a store to memory location [C]. Since memory location [C] was cached in multiple caches (cache 1 and cache 4), the store operation to memory location [C] had to invalidate other cached copies of memory location [C] (including the cached copy in cache 4). However, due to the bug, the invalidation message received by cache 4 was delayed by 5 clock cycles. Before the invalidation occurred, processor core 4 loaded from memory location [C]. Since the cached copy of memory location [C] in cache 4 was still marked as valid, it loaded a stale copy (which contained the wrong value at that point). Then, millions of clock cycles later, processor core 4 used the wrong value of memory location [C] in code that performed locking, resulting in a deadlock.

When such a deadlock is detected (e.g., by using a timeout), the bug must be localized by identifying the bug trace and the component where the bug is located. Since it is not known *a priori* when the bug was activated or when the system deadlocked, it can be very difficult to obtain the bug trace. Additionally, the bug trace can be extremely long due to the long error detection latency, containing extraneous instructions that are not needed for activating or detecting the bug. As discussed above, such bugs are extremely challenging to localize using approaches such as trace buffers, failure reproduction, simulation, or traditional formal methods.

As shown in Sec. IV, Symbolic QED correctly localizes this bug to cache 4 and produces a bug trace that is only 3 instructions long. Symbolic QED takes only 2.5 hours to automatically localize this bug without requiring any failure reproduction, or any additional hardware. This is possible because Symbolic QED uses bounded model checking (BMC), which finds the minimal bug trace, if one exists [27] (details in Sec. III). Additionally, Symbolic QED employs special "design reduction" techniques to effectively handle large multi-core SoC designs such as the OpenSPARC T2 SoC (details in Sec. III). In contrast, traditional post-silicon bug localization approaches would likely require manual effort, additional hardware (e.g., trace buffers), or both, and could take days or weeks (or even months). Furthermore, the bug traces found by traditional post-silicon techniques can be significantly longer than those found by Symbolic QED (empirically demonstrated in Sec. IV).

While the motivation and main focus of this paper is post-silicon bug localization, the Symbolic QED technique can also be used for logic bug detection and localization during pre-silicon verification, as well as emulation-based verification, **without** significant changes.

The rest of this paper is organized as follows. Sec. II provides an overview of the previously-published Quick Error Detection (QED) technique. Sec. III presents the Symbolic QED technique. Results are presented in Sec. IV, followed by related work in Sec. V. We conclude in Sec. VI, with supplemental materials in the appendices.

## II. Background: Quick Error Detection (QED)

QED tests have been demonstrated to be highly effective for quickly detecting logic and electrical bugs inside processor cores, uncore components, accelerators, and components related to power-management features [12-15, 28]. The software-only QED technique automatically transforms post-silicon validation tests (*original tests*) into new *QED tests* using various QED transformations, e.g., Error Detection using Duplicated Instructions for Validation (*EDDI-V*), Proactive Load and Check (*PLC*), Control Flow Checking using Software Signatures for Validation (CFCSS-V), and Control Flow Tracking using Software Signatures for Validation (CFTSS-V).

### A. *EDDI-V*

EDDI-V [12, 14] targets bugs inside processor cores by frequently checking the results of *original* instructions against the results of *duplicated* instructions created by EDDI-V. First, the registers and memory space are divided into two halves,[2] one for the original instructions and one for the duplicated instructions. Next, corresponding registers and memory locations for the original and the duplicated instructions are initialized to the same values. Then, for every load, store, arithmetic, logical, shift, or move instruction in the original test, EDDI-V creates a corresponding duplicate instruction that performs the same operation, but on the registers and memory reserved for the duplicate instructions. The duplicated instructions execute in the same order as the original instructions.

[1] Uncore components refer to components in an SoC that are neither processor cores nor co-processors. Examples include interconnect fabrics, and cache / memory controllers.

[2] For EDDI-V, if it is not possible to divide the registers into two halves (e.g., if the original test needs to use all of the available registers), we can use memory to store the register values. The details are in [Lin 14].

The EDDI-V transformation also inserts periodic check instructions (referred to as *Normal checks* in this paper) that compare the results of the original instructions against those of the duplicated instructions. For every duplicated load instruction, an additional *Load check* instruction is inserted immediately after (before the loaded values are used by any other instructions) to check that the value loaded by the original instruction matches the value loaded by the corresponding duplicated instruction. Similarly, for store instructions, a *Store check* instruction is inserted immediately before the original store instruction to check that the value about to be stored by the original instruction matches the value about to be stored by the duplicated instruction. Each check instruction is of the form:

CMP Ra, Ra' ,

where *Ra* and *Ra'* are the original and (corresponding) duplicate registers, respectively. A mismatch in any check instruction indicates an error. To minimize any intrusiveness [13, 14, 15] that might prevent bug detection by QED, insertion of the duplicated instructions and the check instructions is controlled by the parameters *Inst_min* and *Inst_max*, the minimum (maximum) number of original instructions that must (can) execute before any duplicated or check instructions.

### B. PLC

PLC targets bugs inside uncore components by frequently and proactively performing loads from memory (through those uncore components) and checking the values loaded. Starting with an EDDI-V-transformed QED test, PLC inserts Proactive Load and Check operations (*PLC operations*) throughout the transformed test, which runs on all cores and all threads. In Fig. 3(a), a segment of code has been transformed with PLC operations, with the PLC operation detailed in Fig. 3(b). Each PLC operation checks the values in memory for a selected set of variables (*PLC list*). For each selected variable, a PLC operation loads the value from the memory reserved for original instructions (address *A*) and the value from the corresponding memory reserved for duplicated instructions (address *A'*). Any mismatch during the PLC check (*CMP Rt, Rt'*) indicates an error. Note that locks must be used if the variable is shared between multiple cores / threads or if there are sources of non-determinism in the system (e.g., interrupts, I/O, or OS functionalities such as context switches). Several PLC strategies are discussed in [13, 14, 15].

### C. CFCSS-V and CFTSS-V

Control Flow Checking using Software Signatures for Validation (CFCSS-V) and Control Flow Tracking using Software Signatures for Validation (CFTSS-V) are two QED transformations that target bugs that affect a processor core's control flow [14].

```
//initialization                   //initialization
R1  = 1                            R1  = 1   R17  = 1
R2  = 2                            R2  = 2   R18  = 2
R3  = 3                            R3  = 3   R19  = 3
R4  = 4                            R4  = 4   R20  = 4
R5  = 5                            R5  = 5   R21  = 5
R6  = 6                  ➔         R6  = 6   R22  = 6
//code                             //code
R1 = R2 + R3                       R1  = R2  + R3
R4 = R5 - R6                       R4  = R5  - R6
R4 = R1 - R4                       R4  = R1  - R4
B label                            R17 = R18 + R19
                                   R20 = R21 - R22
                                   R20 = R17 - R20
                                   CMP R4, R20
                                   BNE ERROR_DETECTED
                                   B label
```

**Figure 2**. EDDI-V transformation, with *Inst_min = Inst_max = 3.*

```
Transformed Code          | PLC Operation
   ...                    |
<PLC Operation>           |
R1  = R2  + R3            | for <A,A'> in PLC_list do
R4  = R5  - R6            |   LOCK(A)
R17 = R18 + R19           |   LOCK(A')
R20 = R21 - R22           |   Rt  = LOAD(A)
<PLC Operation>           |   Rt' = LOAD(A')
R7  = R1  - R4            |   UNLOCK(A')
R9  = R7  * R8            |   UNLOCK(A')
R23 = R17 - R20           |   CMP Rt, Rt'
R25 = R23 * R24           |   BNE ERROR_DETECTED
<PLC Operation>           | end for
   ...                    |
```

**Figure 3.** PLC example with *Inst_min = Inst_max = 4.*

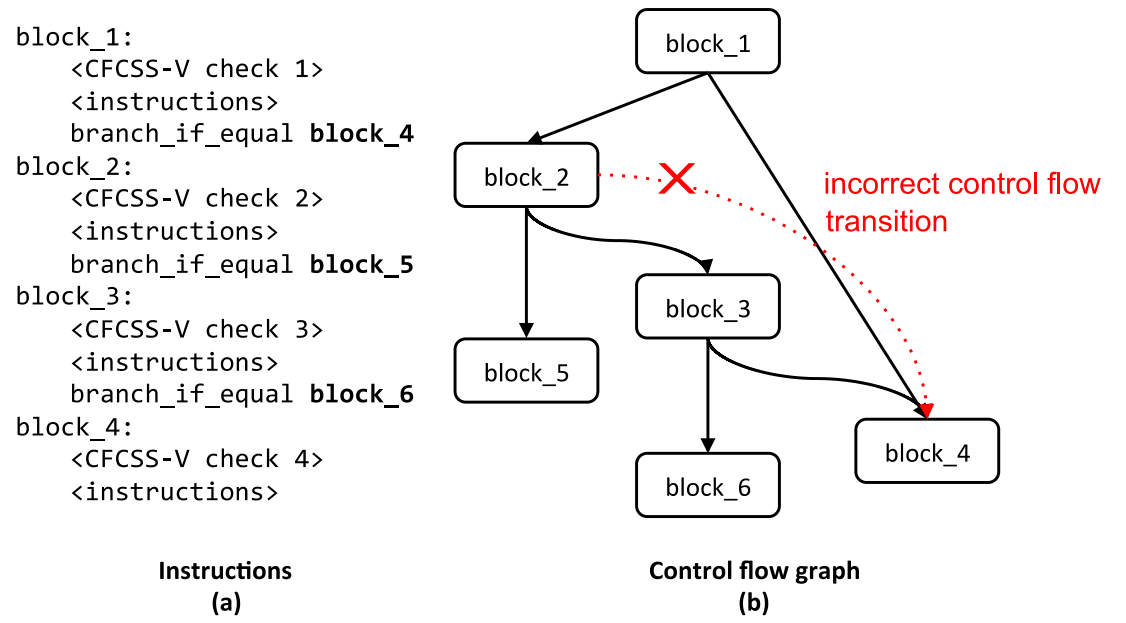

**Figure 4.** (a) An example sequence of instructions, and (b) the corresponding control flow graph. The dashed arrow represents an incorrect control flow transition from block_2 to block_4.

```
if ((signature == block_1_sigature) or (signature == block_3_signature))
then
  signature = block_4_signature // update signature
else
  ERROR DETECTED
end if
```

**Figure 5.** Pseudo code for <CFCSS-V check 4>.

```
ST [TEMP_VARIABLE], R1
LI R1, SOFTWARE_SIGNATURE
ST [CFTSS_V_SIGNATURE], R1
LD R1, [TEMP_VARIABLE]
```

**Figure 6.** Pseudo assembly code for the CFTSS-V operation inserted at the beginning of each "block of instructions".

Consider a sequence of instructions shown in Fig. 4(a), and the corresponding control flow graph (Fig. 4(b)). A CFCSS-V check block is inserted at the beginning of each block of instructions. The pseudo code for <CFCSS-V check 4> is shown in Fig. 5. In Fig. 5, block_1_signature, block_3_signature, and block_4_signature are the unique software signatures for each block. When the control flow transitions from one block to another, the CFCSS-V check block is executed first (before any other instructions in the block). In the case of block_4, <CFCSS-V check 4> checks to see if the signature variable (which contains the software signature of the last block of instructions executed) is equal to either block_1_signature or block_3_signature (since these are the only two blocks that may execute immediately before block_4).

CFTSS-V is a variant of CFCSS-V that tracks the execution of instructions using special software signatures inserted into the test code, but does not perform control flow checking. We first declare a global variable CFTSS_V_SIGNATURE that holds the current runtime signatures of the program. Next, the instructions in the test are divided into "blocks of instructions" as determined by the QED transformation parameters *Inst_min* and *Inst_max*. A unique integer (i.e., software signature) is assigned to each "block of instructions" [29-31]. Fig. 6 lists the pseudo assembly code for the CFTSS-V operation. In Fig. 6, [TEMP_VARIABLE] is a designated memory location used to store the saved value of R1. It ensures the CFTSS-V transformation does not alter the value of the registers in the original test. SOFTWARE_SIGNATURE is the unique software signature assigned to the "block of instructions" using the algorithm found in [30], and [CFTSS_V_SIGNATURE] is the designated memory location to hold the current runtime signature of the test. When a failure occurs (e.g., livelock or deadlock) the content of [CFTSS_V_SIGNATURE] is used to find the last "block of instructions" executed by the processor core before the failure.

## III. SYMBOLIC QED

Symbolic QED detects and localizes bugs, and produces short bug traces consisting of only a few instructions (often less than 10) automatically. Within the space of QED-compatible bug traces (explained below), the traces produced by Symbolic QED are *minimal*, meaning no shorter bug traces exist. These short bug traces make bugs easier to understand and fix.

The Symbolic QED approach presented in this paper relies on bounded model checking (*BMC*), a technique used in formal verification. Given a model of a system (e.g., the RTL) and a property to be checked (e.g., a check inserted by QED), the system is formally analyzed to see if the property can be violated in a bounded number of steps (clock cycles). If so, a *counter-example* (a concrete trace violating the property, i.e., a bug trace) is produced. BMC guarantees that if a counter-example is found, it is a minimal-length counter-example [27]. We first review three challenges associated with using BMC for post-silicon bug localization: 1) BMC needs a property to check. Since the bugs are not known *a priori*, it is difficult to craft such properties (and avoid false positives); 2) Large design sizes limit the effectiveness of BMC. If a design is too large, a typical BMC tool will not even be able to load the design (see Sec. IV). Even if a large design can be loaded, running BMC on it is likely to be very slow; and, 3) the performance of BMC techniques is affected by the number of cycles required to trigger and observe a bug. As the number of cycles increases, BMC performance slows down, especially for large designs. Thus, unless a short counter-example exists, BMC will take too long or will be unable to find it.

We address challenge (2) in Sec. III.E. Here, we focus on challenges (1) and (3). The key idea is to create a BMC problem that searches through ***all possible*** *QED tests* (for a given set of QED transformations). As shown in [12, 13, 14, 15], QED tests are excellent for detecting a wide variety of bugs; hence, we use QED checks (i.e., Normal checks, Load checks, Store checks, and PLC checks) as the properties, thus addressing challenge (1). QED tests are also designed to detect errors quickly. By searching all possible *QED tests* using the minimality guarantees of BMC, it is usually possible to find a very short trace triggering the bug, addressing challenge (3). The details of Symbolic QED using the EDDI-V and PLC QED tests are explained in the following subsections, followed by an extension of the methodology to CFCSS-V and CFTSS-V.

### A. Solving for QED-Compatible Bug Traces Using BMC

For EDDI-V and PLC transformations, QED tests provide very succinct properties to check using check instructions of the form:

CMP Ra, Ra'.

For PLC checks and Load checks, *Ra* and *Ra'* hold values loaded from uncore components; for Normal checks and Store checks, *Ra* and *Ra'* hold the results of computations executed on the cores. An error is detected when the two registers are not equal. Thus, we use BMC to find counter-examples to properties of the form:

Ra==Ra',

where *Ra* is an original register and *Ra'* is the corresponding duplicated register. However, without additional constraints, the BMC engine will find trivial counter-examples that do not correspond to real bugs. For example, the instruction sequence {MOV R1←1, MOV R17 ←2, CMP R1, R17} results in R1≠R17; the inequality is not caused by a bug. In order to avoid such situations, we require that counter-examples must be *QED-compatible*. We define a *QED-compatible* bug trace as a sequence of inputs with the following properties:

1. Inputs must be valid instructions. Specifications of valid instructions can be directly obtained from the Instruction Set Architecture (ISA) of the processor cores.
2. The registers and memory space are divided into two halves: one for "original" instructions and one for "duplicated" instructions. For every instruction (excluding control-flow changing instructions) that operates on the registers and memory space allocated for the original instructions, there exists a corresponding duplicated instruction that performs the same operation, but operates on the registers and memory space allocated for the duplicated instructions.
3. The sequence of original instructions and the sequence of duplicated instructions must execute in the same order.
4. The comparison (i.e., the property checked by the BMC tool) between an original register *R* and its corresponding register *R'* occurs only if the original and its corresponding duplicate instructions have both been executed.

To find a bug trace using BMC, Symbolic QED requires three inputs described in the next sections: the design RTL with a QED module (III.B), an initial state (III.C) and a QED-based property to check (III.D and III.E).

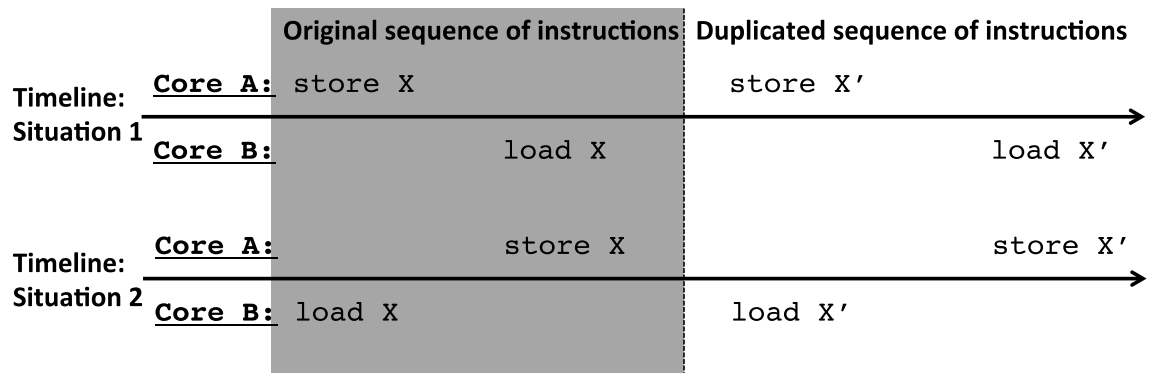

**Figure 7.** A timeline illustrating situations where PLC on variable X does not result in false fail.

### B. QED Module

Ensuring that only QED-compatible bug traces are considered by BMC requires constraining the inputs to the design. We accomplish this by adding a new QED module to the fetch stage of *each* processor core during BMC. **The QED module is only used within the BMC tool and is not added to the manufactured IC**; i.e., there is no performance/area/power overhead. The QED module only needs to be designed once for a given ISA, and made available as a "library component" for use during validation. The design of a QED module is simple, and can be tested in only a few minutes (see Sec. IV). Note that, although the QED module is added to processor cores, Symbolic QED is effective not only for bugs inside processor cores, but also for bugs in uncore components, as well as bugs related to power-management features (as demonstrated in Sec. IV).

The *QED module* automatically transforms a sequence of original instructions into a QED-compatible sequence. Any control-flow altering instruction determines the end of the "sequence of original instructions."[3] The QED module only requires that this sequence is made up of valid instructions and that they read from or write to only the registers and memory allocated for the original instructions (conditions that can be specified directly to the BMC tool). The sequence of original instructions is first executed unmodified (up to but not including the control-flow instruction), and the instructions are committed. Then, it is executed a second time, but instead of using the original registers and memory, the instructions are modified to use the registers and memory allocated for the duplicated instructions. Since duplication is triggered only by a control-flow instruction, the QED module does not use a fixed value for *Inst_min* and *Inst_max*. Instead, (by design) the BMC tool considers counter-examples (in this case, sequences of original instructions) starting with smaller sequences and then moving to longer sequences [27]. This allows the BMC tool to **implicitly** (and **simultaneously**) search through a wide variety of instruction sequences of increasing lengths in order to find a bug trace. After the second execution, a signal is asserted to indicate that the original and corresponding duplicated registers should contain the same values under bug-free situations, i.e., the BMC tool should check the property *Ra == Ra'*.

Note that, because the BMC tool can choose a wide variety of instructions as input to the QED module (including loads and stores), it can effectively create checks that could be generated by a QED transformation, including Normal, Load, Store, and PLC checks. Also note that a PLC check generated by the QED module does not require locks. To see why locks are not needed, we can consider PLC on a variable X as composed of 4 events:

1) Store to the variable X by processor core A,

2) Load from the variable X by a processor core B,

3) Store to the corresponding duplicated variable X' by processor core A, and,

4) Load from the corresponding duplicated variable X' by processor core B.

Note that, multiple processor cores may load from X and X'. In order to guarantee there are no false fails, we need to ensure either one of two situations occur:

**Situation 1:** If event 1 occurs *before* event 2, then event 3 must occur *before* event 4;

**Situation 2:** If event 1 occurs *after* event 2, then event 3 must occur *after* event 4.

[3] One could alternatively use a pseudo-instruction "QED" to trigger instruction duplication; the processor would treat this instruction as a NOP. This would allow the QED module to create sequences that would not be possible otherwise (e.g., an odd number of instructions between two control-flow altering instructions, such as {BRANCH; ADD; BRANCH}).

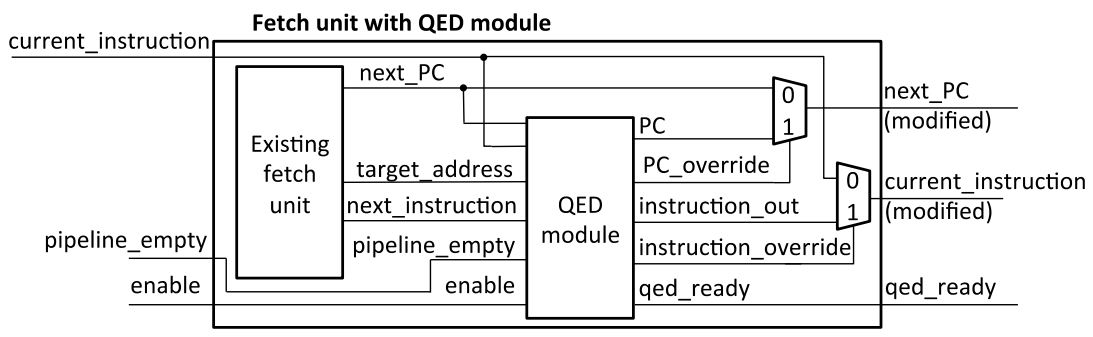

**Figure 8.** The QED module interface.

This means that event 3 and event 4 occur in the same order as that of event 1 and event 2. Figure 7 illustrates these two situations.

Symbolic QED guarantees that only situation 1 or situation 2 will occur. To see why, first, we note that the QED module guarantees that the duplicated sequence of instructions always executes in the same order as that of the original sequence of instructions[4]. Therefore, the only way that event 3 and event 4 (i.e., store to and load from duplicated variable X') will occur in a different order than event 1 and event 2 is if there are delays in executing the duplicated sequence of instructions (by either processor core A or B). For example, in situation 1, if processor core A delays executing the store to the duplicated variable X' until after the load from duplicated variable X' on processor core B. However, there cannot be any delays because:

1) All of the processor cores start executing the duplicated sequence of instructions at the same clock cycle. This is because we wait until the original instructions are committed before executing the duplicated instructions on the same clock cycle. Since the original instructions have committed, no instructions remain in the pipeline that would delay the duplicated instructions, and therefore, the duplicated instructions are guaranteed to start executing on the same clock cycle.

2) The execution of the instructions is deterministic. Therefore, all instructions will complete in a deterministic amount of time (e.g., there are no delays due to context switches).

As a result, Symbolic QED guarantees that PLC will not result in false fails even without locks. This is effective even if there are multiple processor cores loading from variables X and X'. This is because we ensure that all processor cores start executing the duplicated sequence of instructions on the same clock cycle. As a result, the order in which event 3 and event 4 occur will always be the same as the order in which event 1 and event 2 occurred.

Figure 8 shows how the QED module integrates with the fetch unit. The pseudo code of the QED module is shown in Fig. 9. The inputs to the QED module are: 1) *enable*, which disables the QED module if 0 (this signal can be set by the validation engineers to disable the QED module); 2) *current_instruction*, which is the current instruction to be executed in the pipeline of the processor core; 3) *next_instruction*, which is the next sequential instruction after *current_instruction* (i.e., the instruction to be fetched by the fetch unit after *current_instruction*); 4) *next_PC*, which is the PC corresponding to *next_instruction*; 5) *target_address*, which is equivalent to *next_PC* unless the current instruction is a control-flow instruction, in which case it is the control-flow instruction's target address; and 6) *pipeline_empty*, which is a signal that is true if and only if there are no instructions in the pipelines of any of the processor cores and all executed instructions on all cores have been committed (i.e., the results written to registers or to memory).

```
INPUT: enable, current_instruction, next_instruction, next_PC, target_address,
        pipeline_empty
OUTPUT: PC, PC_override, instruction_out, instruction_override, qed_ready
// initialization
mode ← ORIG; // "mode" is shared by all QED modules
rewind_address ← PC obtained from initial state (Sec. III.C);
qed_ready ← false;  PC_override_i ← 0;        instruction_override_i ← 0;
// end initialization
PC_override ← enable ? PC_override_i : 0;
instruction_override ← enable ? instruction_override_i : 0;
if mode == CHECK then
  mode ← ORIG;                              qed_ready ← true;
  PC ← target_address;                      PC_override_i ← 1;
  rewind_address ← target_address;
end if
if mode == ORIG, then
  qed_ready ← false; instruction_override_i ← 0; PC_override_i ← 0;
  if is_control_flow_instruction(next_instruction) then
    mode ← WAIT1; // all QED modules go to "WAIT1" when any QED
                  // module gets a control-flow instruction
  end if
end if
if mode == WAIT1 then // wait until pipelines of all processor cores are empty
  mode ← pipeline_empty ? DUP : WAIT1;      qed_ready ← false;
  instruction_out ← NOP;                     instruction_override_i ← 1;
  PC ← rewind_address;                       PC_override_i ← 1;
end if
if mode == DUP then
  qed_ready ← false;  rewind_address ← next_PC;  PC_override_i ← 0;
  if is_control_flow_instruction(next_instruction) then
    mode ← WAIT2; // all QED modules go to "WAIT2" when any QED
                  // module gets a control-flow instruction
  end if
  instruction_out ← create_duplicated_version(current_instruction);
  instruction_override_i ← 1;
end if
if mode== WAIT2 then // wait until pipelines of all processor cores are empty
  mode ← pipeline_empty ? CHECK : WAIT2;  qed_ready ← false;
  instruction_out ← NOP;                     instruction_override_i ← 1;
  PC ← rewind_address;                       PC_override_i ← 1;
end if
```

**Figure 9.** Pseudo code for QED module.

The outputs from the QED module are: 1) *PC*, which is used to override the value of *next_PC*; 2) *PC_override*, which determines if the processor core should use the *PC* from the QED module or *next_PC* from the fetch unit; 3) *instruction_out*, which is used to override the value of *current_instruction*; 4) *instruction_override*, which determines whether the processor core should use the modified instruction (*instruction_out*) from the QED module or *current_instruction*; and 5) *qed_ready*, which signals when both original and duplicated registers should have the same values (under bug-free conditions). *qed_ready* is false initially; it is only set to true when both original and duplicated instructions have committed.

The QED module has internal variables: 1) *mode*, which tracks if the processor core is executing original instructions (*ORIG*), duplicated instructions (*DUP*), in a wait mode (*WAIT1* or *WAIT2*), or if the BMC tool should do a check (*CHECK*). This variable is shared by all of the QED modules in the design so that they are always in the same mode; 2) *rewind_address*, which holds the address of the first instruction in the sequence of original instructions, (initialized to PC obtained from the initial state in Sec. III.C); 3) *PC_override_i* and 4) *instruction_override_i*, which are internal versions of *PC_override* and *instruction_override* (the only difference is that when the *enable* is set to 0, then both *PC_override* and *instruction_override* are also set to 0, disabling the QED module). The QED modules start in *ORIG* mode. When *next_instruction* is a control-flow altering instruction, all QED modules go to *WAIT1*. In *WAIT1*, *PC* is set to *rewind_address*, and *PC_override_i* is set to 1 (if *enable* is 1, *PC_override* is also set to 1). The QED module also outputs NOP on *instruction_out* and sets *instruction_override_i* to 1 (if *enable* is 1, *instruction_override* is also set to 1). The QED modules stay in *WAIT1* until all of the original instructions have committed (when *pipeline_empty* becomes true, i.e., all processor core pipelines are empty). Then, all QED modules switch to *DUP*, and each processor core then re-executes instructions starting from the address stored in *rewind_address*. In *DUP*, the duplicated instruction is produced on *instruction_out*, and *instruction_override_i* is set to 1, so the core executes the duplicated instruction instead of the original instruction from the fetch unit. In *DUP*, *rewind_address* is constantly updated to *next_PC*.

Then, when *next_instruction* is a control-flow altering instruction, all QED modules switch to *WAIT2* and stay in *WAIT2* until the duplicated instructions on all processor cores have committed (the *pipeline_empty* signal becomes true, i.e., the pipelines of all processor cores are empty). In *WAIT2*, *PC* is set to *rewind_address*. The QED module also outputs NOP on *instruction_out* and sets *instruction_override_i* to 1 (if *enable* is 1, *instruction_override* is also set to 1). After the instructions have committed, the original and corresponding duplicated registers should be equal. Then, the QED modules switch to *CHECK*. In *CHECK*, *qed_ready* is set to true. Each QED module also updates *rewind_address* to *target_address* (i.e., the address of the next instruction to execute) and sets *PC* to *target_address* and *PC_override_i* to 1. After *CHECK*, the QED modules return to *ORIG*.

An example of the transformation performed by the QED module is shown in Fig. 10. Note that, LOAD(A) is transformed into LOAD(A') during the second execution. Thus, comparing the registers (using the BMC tool) is equivalent to a PLC check on variables A and A'. There are 4 events here: (1) store to A by core 1, (2) load from A by core 2, (3) store to A' by core 1, and (4) load from A' by core 2.

[4] This is because (as explained later) for the duplicated execution, the QED module modifies the operands of the original sequence of instructions such that they operate on registers and memory reserved for the duplicated sequence of instructions (the order of the instructions is preserved).

| Core 1 | Core 2 |
|---|---|
| A  = STORE(R1) | R2 = R3 - R4 |
| R2 = R3 + R4 | R1 = LOAD(A) |
| R5 = LOAD(A) | R5 = LOAD(B) |
| BRANCH label | |

(a)

| Core 1 | Core 2 |
|---|---|
| A  = STORE(R1) | R2 = R3 - R4 |
| R2 = R3 + R4 | R1 = LOAD(A)   // PLC load |
| R5 = LOAD(A)  // PLC load | R5 = LOAD(B) |
| **A' = STORE(R17)** | **R18 = R19 - R20** |
| **R18= R19 + R20** | **R17 = LOAD(A') // PLC Load** |
| **R21= LOAD(A') // PLC load** | **R21 = LOAD(B')** |
| BRANCH label | |

(b)

**Figure 10.** Example of QED transformation by the QED module. (a) A sequence of original instructions on core 1 and core 2, and (b) the actual transformed instructions executed by the cores.

As explained earlier in this Section, to avoid false fails without using locks, the QED module ensures that the order of (3) and (4) is the same as the order of (1) and (2), even if multiple cores load from A and A'. Because the BMC tool can choose a wide variety of instructions for the original sequence of instructions, this does not significantly affect the ability of Symbolic QED to activate and find bugs in general (which is empirically demonstrated in Sec. IV). However, in future work, one may want to allow the processor cores to have more freedom when executing the duplicated instructions; in that case, locks may be necessary. Memory initialization is discussed in Sec. III.C.

### *C. Initial State*

The approach outlined above ensures that only QED-compatible traces are considered by BMC. However, the initial state for the BMC run must be a *QED-consistent* state, in which the value of each register (in the processor core) and memory location allocated for original instructions must match the corresponding register or memory location for duplicated instructions. This is to ensure that no false counter-examples are generated. One approach would be to start the processor from its reset state. However, the reset state may not be QED-consistent (or it may be difficult to confirm whether it is). Some designs also go through a reset sequence that may span several clock cycles, making the BMC problem more difficult. For example, for OpenSPARC T2, only one processor core is active after a reset, and the system executes a sequence of initialization instructions (approximately 600 clock cycles long) to activate other processor cores in the system.

It is advantageous to start from a QED-consistent state after the system has executed the reset sequence (if any) to improve the runtime of BMC (also demonstrated by results in Sec. IV). A simple way to obtain a QED-consistent state is to run “some” QED test (independent of specific tests for bug detection and debug) in simulation and to stop immediately after QED checks have compared all of the register and memory values (this ensures that each “original” register or memory location has the same value as its corresponding “duplicate” register or memory location). This can be accomplished with a simple (short) test that just writes to the original and corresponding duplicated registers and memory locations and checks them to ensure that they are in a QED-consistent state. The register values (including the *PC* and *next_PC* from Sec. III.B) and memory values are read out of the simulator and then used to set the register values, *PC*, *next_PC*, and memory values of the design when preparing to run BMC. If the design contains multiple processor cores, the processor cores can be simulated together. Alternatively, each core can be simulated independently and the results merged together to set up the BMC run. In this case, some care must be taken to ensure that the values in shared memory locations are the same at the end of each simulation (e.g. by running the same test on each core). One can obtain these values using ultra-fast simulators (at a higher level of abstraction than RTL) that can simulate large designs with thousands of processor cores [32]. Thus, this initialization step does not affect the scalability of Symbolic QED.

It is possible that some bugs may not be detected from a generic QED-consistent initial state as they require a very long activation sequence. We target these bugs during post-silicon validation by leveraging QED tests, as presented in Sec. IV.D.

### *D. Finding Counter-Examples using BMC*

After inserting the QED module and setting the initial state, we use BMC to find a counter-example to the property:

$$qed_ready \rightarrow \bigwedge_{a\in\{0..\frac{n}{2}-1\}} Ra == Ra',$$

where $n$ is the number of registers defined by the ISA. Here (for $a \in \{0..n/2-1\}$), $Ra$ and $Ra'$ correspond to registers allocated for original instructions and duplicated instructions respectively. As mentioned above (e.g. Fig. 10), because we allow the instructions chosen by BMC to include load and store instructions, our approach can generate PLC checks, and can thus activate and detect bugs in uncore components as well as those in processor cores.

### *E. Symbolic QED for CFCSS-V and CFTSS-V*

As detailed in Sec. II.C, the CFCSS-V QED transformation assigns unique software signatures to instructions or blocks of instructions. It then checks if each transition between instructions or blocks of instructions is valid. To implement CFCSS-V with Symbolic QED, a separate CFCSS-V module is added to the design to decode control flow instructions and determine the valid next instructions. Symbolic QED uses the *memory address* of an instruction (i.e., the memory location where the instruction resides) as its unique software signature. The CFCSS-V module decodes the memory addresses of the target instructions and stores them. The module then compares these memory addresses against the memory address of the next instruction to execute to determine if it is a valid instruction. The BMC tool is then asked to find a counter-example instruction trace that executes an invalid instruction.

To implement CFCSS-V in a BMC tool, the property to be checked by the tool is:

(currently_executing_PC==ADDRESS_branch)
or
(currently_executing_PC==ADDRESS_next),

where *currently_executing_PC* is the memory address (i.e., the content of PC or program counter) of the instruction that the processor core is currently executing. For in-order processors, this corresponds to the PC of the instruction in the execution stage; for processors with out-of-order or speculative execution, this is the PC of the most recently committed instruction. The CFCSS-V module reads this instruction to decode its type and stores the appropriate *ADDRESS_branch* and *ADDRESS_next* values for control flow instructions. The instruction type (i.e., conditional branch, branch always, or non-control-flow altering instruction) is used as a select signal for a MUX that is used to determine what values (e.g., address of first instruction in the taken branch, address of first instruction in non-taken branch, or an increment of the PC) are used for *ADDRESS_branch* and *ADDRESS_next*. If the last executed instruction is a conditional branch instruction (determined by decoding the instruction), the *ADDRESS_branch* is the address of the first instruction in the taken branch (also obtained by decoding the instruction), and *ADDRESS_next* is the address of the first instruction in the not taken branch (usually an increment of the PC). If the last executed instruction is a non-conditional branch instruction (branch always) both *ADDRESS_branch* and *ADDRESS_next* are assigned to the target address of the branch instruction. If the last executed instruction does not alter control flow (e.g., ALU instructions, load / store instructions), *ADDRESS_branch* is set to the same value as *ADDRESS_next* (usually it is an increment of the PC).

Unlike EDDI-V and PLC, Symbolic QED for CFCSS-V doesn't insert additional instructions. Hence, the *Inst_min* and *Inst_max* transformation parameters are not relevant. Symbolic QED for CFCSS-V requires the following input constraint:

```
Inputs must be valid instructions (specifications of valid instructions can be directly obtained from the Instruction Set Architecture (ISA) of the processor cores).
```

Since the BMC tool searches through all possible sequences of instructions for the input (including sequences of instructions that contain branch instructions), if one or more branch instructions are required to activate and detect a bug, the BMC tool will automatically find a bug trace that contains the necessary branch instruction(s) to violate the property as a counter-example.

CFTSS-V tracks the control flow between blocks of instructions, and is extremely useful for deadlock / livelock situations. For bug localization during post-silicon validation, one may not necessarily need to use Symbolic QED for CFTSS-V. This is demonstrated in [14], where QED family tests (with a range of different values for the *Inst_min* and *Inst_max* QED transformation parameters and different windows for the QED transformations) are sufficient to find a very short bug trace (e.g., 9 instructions in [14]). However, Symbolic QED for the CFTSS-V QED transformation can be used to produce short bug traces during pre-silicon verification. In this context, deadlock is a situation during which signals in the processor core do not change, and the processor core does not make progress [33], so in our case it does not commit any instructions. During livelock, some signals in the processor core do change, but the processor core still does not make progress, and again in our examples it again does not commit any instructions. Note, we do not consider tests that contain self-loops (e.g., an infinite loop) to be livelocks.

For Symbolic QED using CFTSS-V, we use the following property for BMC:

**F** (number of committed instructions == C)

This property states that eventually (i.e., F), the processor core has to commit C number of instructions. Note that the BMC tool would not attempt a trace of N+1 instructions without exhaustively trying all traces of N instructions. Thus, a counter-example to this property corresponds to a situation where the processor core is unable to commit C number of instructions (due to livelocks / deadlocks). The parameter C should be set as large as possible. However, since the design is analyzed in a BMC tool, which can only analyze a limited number of clock cycles (due to the size of the design), C is limited by the number of clock cycles the BMC tool can analyze. For example, if a processor core can only commit 1 instruction every 2 clock cycles and the BMC tool can only analyze 10 clock cycles of the design, it is unreasonable to ask the BMC tool to find a situation where the processor core is unable to commit 1,000 instructions (i.e., C = 1,000). Therefore, C corresponds to the number of instructions that the processor core can guarantee to commit in the given number of clock cycles analyzed by the BMC. The parameter C depends on the BMC tool used and the design being analyzed. For the OpenSPARC T2 processor core design, we determined that that the maximum number for C is 17 instructions for our BMC tool. C is obtained empirically. We start by setting C = 1, and analyze the design (including the QED module) with the CFTSS-V property. If the BMC tool is able to analyze the design (i.e., determine that the design can commit C number of instructions or produce a counter-example indicating that C number of instructions cannot be committed), then we increase C by 1. We keep doing so until the BMC tool is unable to determine if C number of instructions can be committed or produce a counter-example after 50 hours.

To keep track of the number of instructions that have committed, we add a small counter that counts the number of instructions committed (Fig. 11). This counter is only used within the BMC tool and is not added to the manufactured IC. The counter counts up by 1 every time an instruction is committed, which is determined when an instruction commit signal from the commit stage of a processor core is active. The output of the counter corresponds to the number of committed instructions (used by the property check by the BMC tool). The counter is reset to 0 at the start of BMC.

Similar to Symbolic QED for CFCSS-V, we require the following constraints on the inputs.

```
Inputs must be valid instructions (defined again
by the Instruction Set Architecture (ISA) of the
processor cores).
```

Symbolic QED for CFTSS-V may be combined with the Symbolic QED for EDDI-V, PLC and CFCSS-V.

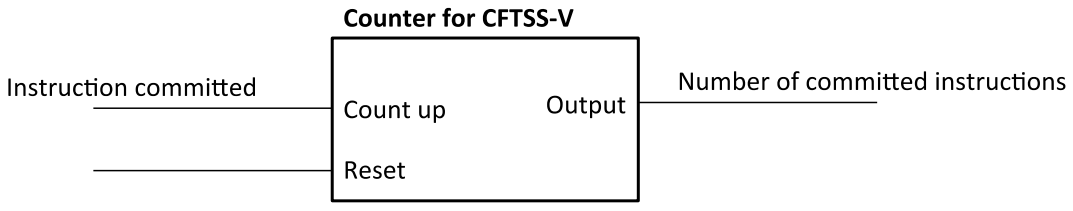

**Figure 11.** The counter for CFTSS-V.

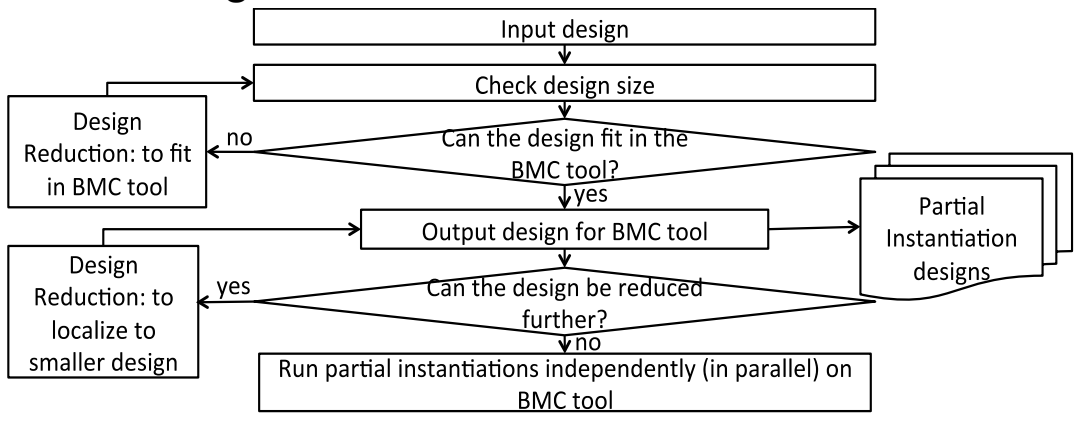

**Figure 12**. The partial instantiation approach for design reduction.

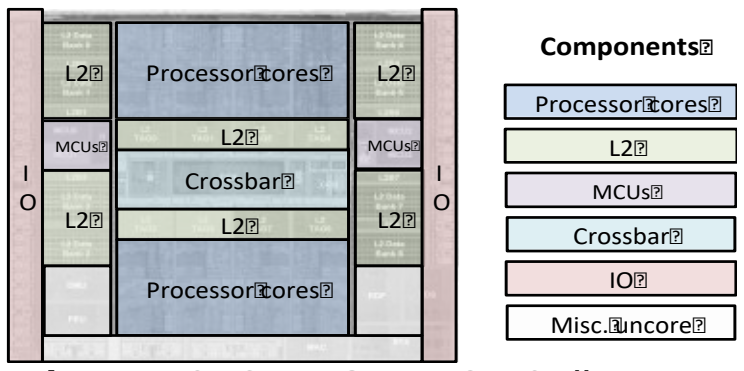

**Figure 13**. OpenSPARC T2 diagram.

## F. *Handling Large Designs*

A major challenge with BMC is handling large designs, which can significantly slow down a BMC tool or even cause it to fail while loading the design. A state-of-the-art commercial BMC tool may not be able to load a complete SoC (e.g., this is the case for OpenSPARC T2). However, it is not necessary to analyze an entire design at once using Symbolic QED. A key property of QED checks is that they are compositional, i.e., they are preserved across partial instances of a design (as long as there is one processor core in that instance). Symbolic QED handles large designs through partial instantiation, as described in the next section, a method that requires no hardware overhead and can be applied to any SoC design.

We also discuss two further techniques (Core vs. Uncore Localization, Change Detectors) for helping handle such large designs if standard QED post-silicon validation tests are run before attempting to localize the bug. These techniques are not required, if Symbolic QED is used independently during pre- or post-silicon design stages.

### *Partial Instantiation*

Partial instantiation works through two design reduction techniques. *Technique 1* takes all components with multiple instances and repeatedly reduces their count by half until there is only 1 left. For example, in a multi-core SoC, the processor cores are removed from the design until there is only 1 processor core left. *Technique 2* removes a module as long as its removal does not divide the design into two disconnected components. For example, if a design has a processor core connected to a cache through a crossbar, the crossbar is not removed (without also removing the cache). This is because if the crossbar is removed, the processor core is disconnected from the cache. *All possible* combinations and repetitions of the two techniques are considered when producing candidates for analysis. Since we find bug traces in the form of instructions that execute on processor cores, each analyzed design must contain at least one processor core. Fig. 12 shows the steps for this approach. Once the full set of simplified (partially instantiated) designs is created, they can be analyzed by the BMC tool independently (in parallel).

Consider the OpenSPARC T2 design with 8 processor cores, 1 crossbar, 8 banks of shared L2 cache, 4 memory controllers, and an I/O controller (Fig. 13). This entire design is too big to be analyzed by the BMC tool, so it is not saved as a partial instance. One possibility is to remove the I/O controller, resulting in 8 processor cores, 1 crossbar, 8 banks of cache, and 4 memory controllers; this is still too big for the BMC tool, and it is not saved as a partial instance. Alternatively, components with multiple instances (e.g., the cores, caches, and memory controllers) can be halved, reducing the design to 4 processor cores, 1 crossbar, 4 banks of cache, 2 memory controllers, and the I/O controller. This still does not fit in the BMC tool, and so again, it is not saved as a partial instance. At this point, we can take either of our two reduced designs as candidates for further reduction. Let us consider the second one. The crossbar is not removed, as it would disconnect the processor cores from the other components. Suppose instead that we apply technique 1 again. This reduces the design to 2 processor cores, 1 crossbar, 2 banks of cache, 1 memory controller, and the I/O controller. This design still does not fit. Next, either the I/O controller or the memory controller can be removed through technique 2. By removing the I/O controller, we are left with 2 processor cores, 1 crossbar, 2 cache banks, and 1 memory controller. This fits in the BMC tool and so the configuration is saved. Alternatively, by removing the memory controller, we are left with 2 processor cores, 1 crossbar, 2 cache banks and the I/O controller, which fits and is saved.

Now, even though at this point we have two candidate configurations for BMC, we continue to apply design reduction techniques to generate more partial instances. The reason for this is for better localization: if BMC can find a bug trace in a smaller configuration, then this indicates that the components removed by the design reduction techniques are not necessary for activating and detecting the bug. Continuing with the reduction, by applying technique 1, the number of cores and caches can be reduced, resulting in 1 processor core, 1 crossbar, 1 bank of cache, 1 memory controller, and the I/O controller. Further reductions result in smaller and smaller subsets of the design, each of which fits in the BMC tool and is saved. When no more reductions are possible (i.e., the design is reduced down to just a single core), all of the saved designs are analyzed independently (in parallel) by the BMC tool.

*Bugs Inside Processor Cores vs. Outside Processor Cores*

The partial instantiation technique can in some cases be improved if Symbolic QED is being used to further localize a bug found using traditional post-silicon QED. If a (standard, not symbolic) QED test fails either a Normal check or a Store check, we can immediately deduce that the bug is inside the processor core where the check failed.[5] This is because, by design, Normal and Store checks catch any incorrect value produced by a processor core *before* it leaves the processor core and propagates to the uncore components or to other processor cores. Thus, we just need to perform BMC on the single processor core where the check failed in order to find a bug trace. If the test fails at a Load check or a PLC check, we cannot immediately infer where the bug is. For these cases, we would use the Partial Instantiation technique to simplify the design to be analyzed by BMC.

*Change Detectors for Design Reduction*

Symbolic QED does not require any additional hardware (e.g., trace buffers) to be added to the design. During post-silicon validation, however, Symbolic QED can be enhanced using a small amount of additional hardware to reduce the size of the design that the BMC tool must analyze. This hardware does require QED post-silicon validation tests to be run first. We introduced small hardware structures, *change detectors*, to monitor signals between hardware blocks and record if changes occur within a sliding window of immediately preceding execution cycles (e.g. the last 1,000 cycles). When a QED test detects an error, if no signal changes are observed at the boundary of an uncore component (during the execution window being analyzed), then we exclude that component as a candidate contributing to the bug. This reduces the number of components for the BMC tool to analyze. Due to the short error detection latencies with QED (typically less than 1,000 cycles), the monitored sliding window can be small, reducing the area impact.

The relative location and schematic for a change detector is shown in Fig. 14. The change detector is composed of a k-bit ripple counter that is initialized to its maximum value of all 1s and is reset to all 0s whenever a change in signal values is detected. It then begins up-counting increments each clock cycle until it is either reset again (to all 0s) because a change is detected, or it reaches and holds its maximum counter value, indicating no change was detected during the prior (2k-1) cycles. Due to the short error detection latencies of QED tests in [13,14], this time period does not need to be greater than a few thousand cycles (k≈10) to capture the bug activation and error propagation. We define this time period as the change window. When a QED test detects an error, the post silicon validation test stops and the change detector counter values are scanned out and saved.

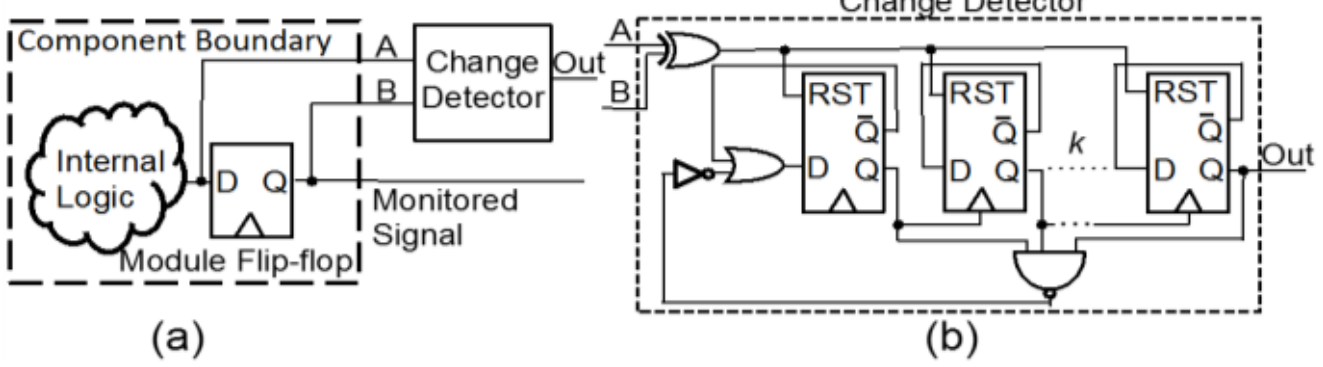

**Figure 14**. **(a)** Insertion of change detector around a flip-flop, and its schematic **(b),** where the *k* scan flip-flops acts as a *k*-bit up counter.

## IV. Results

We demonstrate the effectiveness of Symbolic QED using the OpenSPARC T2 SoC [34] (Fig. 13), which is the open-source version of the UltraSPARC T2, a 500-million-transistor SoC with 8 processor cores (64 hardware threads), private L1 caches, 8 banks of shared L2 cache, 4 memory controllers, a crossbar interconnect, and I/O controllers. We simulated logic bug scenarios from [13-15], which represent a wide variety of "difficult" bug scenarios that occurred in various commercial multicore SoCs. The bug scenarios include bugs in the processor cores, bugs in the uncore components, and bugs related to power-management features.[6] They are considered difficult because they took a long time (days to weeks) to localize. Note that, Symbolic QED does not rely on any information about the specific implementation of OpenSPARC T2, making it applicable to a wide variety of SoCs.

We modified the RTL of the OpenSPARC T2 SoC to incorporate these bug scenarios. For the 80 bug scenarios from [13, 14], we set the bug scenario parameter *X* to 2 clock cycles and bug scenario parameter *Y* to 2 clock cycles. The details of *X* and *Y* are in [14]; note that smaller values for *X* and *Y* imply that the bugs are more difficult to activate and detect. For example, consider the activation criterion 1 from [14]: "two stores within X clock cycles to different cache lines;" and two sequences of instructions: 1) {STORE [a], Rx; STORE [b], Ry} and 2) {STORE [a], Rx; MOV R0, 0; STORE [b], Ry}. While both sequence 1 and sequence 2 will satisfy the activation criterion when X=3 (i.e., two stores within 3 clock cycles to different cache lines), only sequence 1 will satisfy the activation criterion when X=2. For the 12 power management bug scenarios in [15], the activation criterion is set to a sequence of 5 instructions randomly selected from the original test, executed on a designated processor core. This is to emulate a power management controller which puts the system into a power-saving state when it executes a specific sequence of instructions. If a bug is inserted in a component, it is in all instances of that component.

For BMC, we used the Questa Formal tool (version 10.2c_3) from Mentor Graphics on an AMD Opteron 6438 with 128GB of RAM. We added the QED module described in Sec. III.B to the RTL of the fetch unit in the OpenSPARC T2 processor core. The resulting fetch unit with the QED module was tested using Questa to ensure it correctly transforms a sequence of original instructions into a QED-compatible bug trace. The testing process for 50 sequences of original instructions of varying lengths (1 to 10 instructions long) took approximately 1 minute of runtime. Moreover, we simulated all of the bug traces produced by Symbolic QED (which depends on the QED module) to ensure that they indeed activate and detect the corresponding bugs.

### A. *Effectiveness of Symbolic QED for Logic Bug Detection and Localization during Pre-silicon Verification*

The entries in Table 1 are categorized into processor core bugs, uncore bugs (bugs that are inside uncore components as well as in the interface between processor cores and uncore components), and power management bugs. "Bug trace length (instructions)" shows the [minimum, average, maximum] number of instructions in the bug trace. "Bug trace length (cycles)" represents the [minimum, average, maximum] number of clock cycles required to execute the bug trace. The two numbers are different because the number of cycles per instruction (CPI) is not 1 for all instructions (for example, a load or store instruction may take multiple clock cycles to execute). The reported length for bug traces corresponds to the number of instructions in the trace found by the BMC tool (not including duplicated instructions created by the QED modules). For bugs that are only found by executing instructions on multiple processor cores, the number of instructions for each core may be different. For example, one core could have a bug trace that is 3 instructions long, while another core has a bug trace that is 1 instruction long. We report the length of the longest bug trace in such situations (3 in this example), as all cores must completely execute their corresponding instructions to activate and detect the bug (the cores execute instructions in parallel).

We did not include any results from running BMC without Symbolic QED for three reasons: (i) the full design does not load into the BMC tool; (ii) even if it did, we would need properties to check to run BMC, and there is no clear way to create such properties (other than manual creation which would be subjective and extremely time-consuming); (iii) some traditional design-specific assertions may quickly produce shorter counter-examples (vs. Symbolic QED) during pre-silicon verification for specific bugs in stand-alone design blocks targeted by the assertions. However, they can require significant manual effort (e.g., to create them, to model environmental constraints or to filter spurious counter-examples that cannot occur at the full system level) and they may miss bugs that would be caught by Symbolic QED (an example is presented in Appendix B of [35]). Indeed, the Symbolic QED technique for expressing a generic property to check is a key contribution.

[5] The entire test must be transformed by QED for this to work. If some QED checks are left out, then this cannot be guaranteed. For example, if some Normal checks and Store checks are omitted, an error caused by a bug inside the core may propagate to an uncore component.

[6] Bug scenarios are in the appendix. The bug scenarios were simulated by modifying the RTL of the OpenSPARC T2 SoC design so that, for each bug scenario, if the bug activation criterion is satisfied, the bug effect is simulated.

**Table 1.** Symbolic QED results (initialized from a QED-consistent state obtained by running an FFT QED test) for "difficult" logic bug scenarios from [13-15], For bug traces, we report the [min., average, max.] length in instructions and clock cycles, plus the [min., average, max.] BMC runtimes.

| | | Symbolic QED [min., average, max.] |
|---|---|---|
| Processor core only | Bug trace length (instructions) | **[3, 3, 3]** |
| | Bug trace length (clock cycles) | **[13, 15, 16]** |
| | BMC runtime (minutes) | **[22, 46, 90]** |
| | Bugs detected and localized | **100%*** |
| Uncore | Bug trace length (instructions) | **[3, 4, 4]** |
| | Bug trace length (clock cycles) | **[14, 22, 29]** |
| | BMC runtime (minutes) | **[78,164,188]** |
| | Bugs detected and localized | **100%*** |
| Power management | Bug trace length (instructions) | **[5, 5, 5]** |
| | Bug trace length (clock cycles) | **[17, 19, 22]** |
| | BMC runtime (minutes) | **[205,266,333]** |
| | Bugs detected and localized | **100%** |

**Observation 1:** Symbolic QED correctly and automatically detects bugs during pre-silicon verification, and produces short bug traces for all bugs in less than 7 hours. Symbolic QED is effective for large designs such as the OpenSPARC T2, which are challenging when using traditional pre-silicon verification techniques.

Symbolic QED also localized each of the bugs using the partial instantiation technique (Sec. III.F). The BMC tool analyzed the partial instances in parallel. For the OpenSPARC T2, there were 9 parallel BMC runs for each bug; each run corresponded to one of the following partial instances, which are ranked by size in descending order.[7] 1) 2 processor cores, 2 L2 cache banks, and the I/O controller; 2) 2 processor cores, 2 L2 cache banks, and 1 memory controller; 3) 2 processor cores, and 2 L2 cache banks; 4) 1 processor core, 1 L2 cache bank, 1 memory controller, and the I/O controller; 5) 1 processor core, 1 L2 cache bank, and the I/O controller; 6) 1 processor core, 1 L2 cache bank, and 1 memory controller; 7) 1 processor core and the I/O controller; 8) 1 processor core, 1 L2 cache bank; and 9) 1 processor core. Recall that if a bug is in a component, it is in all instances of the component. For these bugs, the BMC runtime reported corresponds to the runtime of the *smallest* partial instance that produced a counter-example. For example, for a given bug, if both partial instances 6 and 8 produced a counter-example, then only the result from partial instance 8 was reported. This example reveals that the additional components in partial instance 6 were not required for activating or detecting the bug. Specifically, for this example, while both partial instances 6 and 8 contain processor cores and caches, partial instance 8 does not have a memory controller. Thus the memory controller was not required to activate and detect the bug. Note that this partial instance provides a small candidate list of components that may contain the bug.

**Observation 2:** Symbolic QED correctly localizes bugs and provides a list of components corresponding to possible bug locations. Figure 15 reports a breakdown of the bugs localized by Symbolic QED, which correctly localized all 92 bugs. Symbolic QED using partial instantiation (Sec. III.F) localized 26.1% the bugs to exactly 1 processor core; for 56.5% of the bugs, Symbolic QED localized the bug to 1 processor core, 1 L2 cache bank and the crossbar that connects the two; and for 17.4% of the bugs, Symbolic QED localized the bug to 2 processor cores, 2 L2 cache banks, and the crossbar that connects the components.

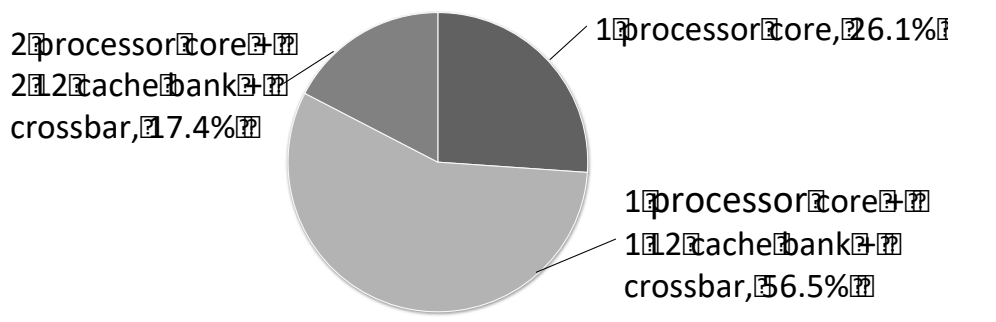

**Figure 15**. Pie chart showing the percentage breakdown (by list of candidate modules) of 92 bugs correctly localized by Symbolic QED.

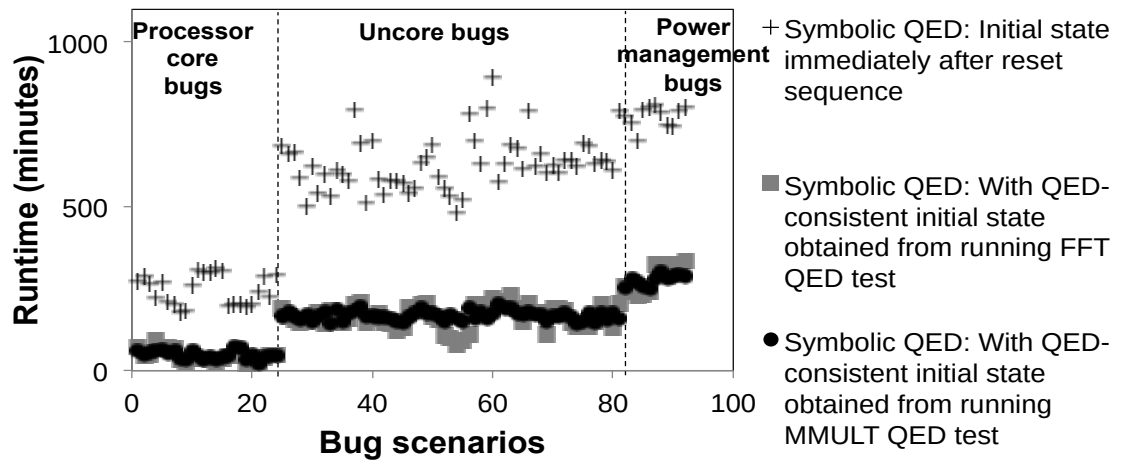

**Figure 16**. The BMC runtimes for Symbolic QED.

The BMC runtimes reported in Table 1 use the QED-consistent initial state constraint discussed in Sec. III.B. In Fig. 16, we report three runtimes for each bug: the runtime when starting from the state immediately after a reset sequence (which is QED-consistent in this case), the runtime when starting from a QED-consistent initial state obtained by running the FFT QED test and seeding BMC with the resulting register and memory values (Sec. III.B), and the runtime when similarly seeding BMC after running MMULT. Results show that using a QED-consistent initial state obtained by running a QED test improves runtimes by up to 5X compared to starting from the state immediately after a reset sequence. No significant differences were observed between the results from using the FFT test and those using the MMULT test.

### *B. Effectiveness of Symbolic QED for Post-silicon Validation*

Here, we consider the scenario when Symbolic QED is not used during pre-silicon verification; and, bugs escape and get detected during post-silicon validation. To compare Symbolic QED against post-silicon validation tests, we used the EDDI-V and the PLC (Sec. II) QED transformations to transform an 8-thread version of the FFT test (from SPLASH-2 [36]) and an in-house 8-thread version of the matrix multiplication test (MMULT) into QED tests to detect bugs. The *Inst_min* and *Inst_max* QED transformation parameters were set to 100, a setting which typically allows bugs to be detected within a few hundred clock cycles (as shown in [12-14]).[8] Trying additional tests (beyond FFT and MMULT) was deemed unnecessary because both tests (after QED transformation) were able to detect all 92 bugs (and Symbolic QED is independent of other tests that detect the bug).

The results are summarized in Table 2. The Original (No QED) column shows results for the original validation tests (FFT or MMULT) using end-result-checks (that check the results of the test vs. pre-computed, known correct results). The QED column shows results from running the same tests after applying QED transformations. Note that, unlike Symbolic QED, both the Original (No QED) and the QED tests (without the techniques discussed in Sec. III.F) are only able to report the existence of a bug; they cannot localize the bug (i.e., determine if the bug is in any of the uncore components, or is caused by interactions between the components); nor can they determine very precisely how the bug is activated. For the post-silicon tests, each entry contains two sets of numbers, corresponding to results obtained from the FFT test (top), and results obtained from the MMULT test (bottom).

**Observation 3:** Symbolic QED **automatically** produces bug traces that are up to 6 orders of magnitude shorter than traditional post-silicon validation tests that rely on end-result-checks, and up to 5 orders of magnitude shorter than QED tests. **Symbolic QED did not require any trace buffers (or any additional hardware) to produce correct bug traces.**

[7] Partial instantiation 1 is the largest that will fit into the BMC tool; all designs also contain the crossbar that connects the components together.

[8] These *Inst_min* and *Inst_max* parameters do not affect the bug traces found by Symbolic QED shown later; they are only used to create the QED tests for detecting bugs.

**Table 2.** Results comparing original tests (No QED) and QED tests for FFT (top values) and MMULT (bottom values) with Symbolic QED (initialized from a QED-consistent state obtained by running an FFT QED test). For bug traces, we report the [minimum, average, maximum] length in instructions and clock cycles. We also report [minimum, average, maximum] BMC runtimes.

| | | Original (No QED) | QED | **Symbolic QED** |
|---|---|---|---|---|
| Processor core only | Bug trace length (instructions) | [643,551k,4.9M]<br>[12k,534k,2.3M] | [324,57k,233k]†<br>[421,67k,321k]† | **[3, 3, 3]** |
| | Bug trace length (clock cycles) | [842,572k,5.1M]<br>[15k,544k,2.5M] | [367,66k,265k] †<br>[522,69k,272k] † | **[13, 15, 16]** |
| | Coverage | 50.0%<br>54.2% | 100%<br>100% | **100%** |
| | BMC runtime (minutes) | N/A | N/A | **[22, 46, 90]** |
| | Bugs localized | 0%<br>0% | 0%<br>0% | **100%*** |
| Uncore | Bug trace length (instructions) | [620,1.6M,9.8M]<br>[1k,536k,2.5M] | [231,59k,232k] †<br>[392,80k,421k] † | **[3, 4, 4]** |
| | Bug trace length (clock cycles) | [722,1.9M,11M]<br>[2k,550k,2.7M] | [292,72k,289k]†<br>[442,95k,435k]† | **[14, 22, 29]** |
| | Coverage | 55.3%<br>57.1% | 100%<br>100% | **100%** |
| | BMC runtime (minutes) | N/A | N/A | **[78,164,188]** |
| | Bugs localized | 0%<br>0% | 0%<br>0% | **100%*** |
| Power management | Bug trace length (instructions) | [1.5k,236k,495k]<br>[963,213k,422k] | [10k,68k,302k]†<br>[1k,47k,134k]† | **[5, 5, 5]** |
| | Bug trace length (clock cycles) | [1.9k,251k,512k]<br>[1.5k,220k,430k] | [13k,75k,319k]†<br>[2k,49k,149k]† | **[17, 19, 22]** |
| | Coverage | 66.7%<br>66.7% | 100%<br>100% | **100%** |
| | BMC runtime (minutes) | N/A | N/A | **[205,266,333]** |
| | Bugs localized | 0%<br>0% | 0%<br>0% | **100%*** |

* Symbolic QED localizes 100% of the bugs *without* using trace buffers.
† If trace buffers are used for QED, then the trace lengths in terms of instructions are: for FFT, [63, 451, 863] for processor core bugs, [29, 487, 832] for uncore bugs, and [42, 297, 742] for power management bugs; and for MMULT, [44, 309, 874] for processor bugs, [32, 502, 884] for uncore bugs, and [67, 392, 742] for power management bugs. The trace lengths in terms of clock cycles are: for FFT, [82, 512, 922] for processor core bugs, [38, 532, 930] for uncore bugs, and [66, 412, 912] for power management bugs; and for MMULT, [69, 420, 921] for processor core bugs, [58, 582, 944] for uncore bugs, and [79, 482, 801] for power management bugs. Other entries remain the same.

These are very difficult bugs that took many days or weeks of (manual) work to localize use traditional approaches (also evident by the long bug traces produced by traditional techniques). Short bug traces make debugging much easier. A more detailed visualization of the trace lengths for each bug scenario is presented in Fig. 17.

With the techniques described in Sec. III.F, standard QED can localize processor core bugs. And in our experiments, as expected, QED tests detected all processor core bugs by either a failing Normal check or a failing Store check, both of which indicate that the bug must be inside a processor core (this was determined solely based on the type of the failing QED check, not because we knew which bugs were being simulated). However, QED fails to provide any localization for uncore bugs. Symbolic QED, if used after QED, can use the information obtained from QED to avoid having to use partial instantiation in the case when QED has already localized the bug to a processor core. In Table 2, "Coverage" is the percentage of the 92 bugs detected. Both Symbolic QED and QED detected all 92 bugs, while the original tests detected only a little more than half of the bugs. This is because original tests (No QED) may not contain the instructions needed to activate a bug, and even if they do, there may not be sufficient checks to detect it.

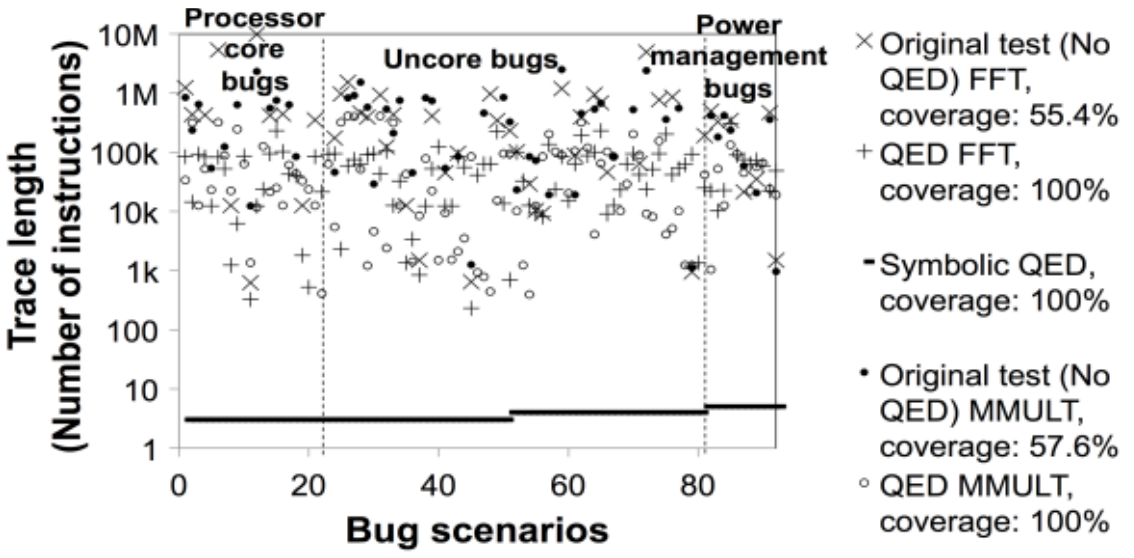

**Figure 17**. Trace length (in terms of number of instructions).

**Table 3**. Results comparing original tests (No QED) and QED family tests (CFCSS-V) for FFT (top values) and MMULT (bottom values), and Symbolic QED for CFCSS-V (initialized from a QED-consistent state obtained by running an FFT QED test). For bug traces, we report the [minimum, average, maximum] length in instructions and clock cycles. We also report [minimum, average, maximum] BMC runtimes in minutes.

| | | Original (No QED) | QED (CFCSS-V) | **Symbolic QED (CFCSS-V)** |
|---|---|---|---|---|
| **Control flow** | Bug trace length (instructions) | [643,321k,1.2M]<br>[12k,319k,1.1M] | [14,39k,91k]†<br>[14,38k,66k]† | **[3, 3, 4]** |
| | Bug trace length (clock cycles) | [781,620k,2.2M]<br>[14k,619k,2.1M] | [22,37k,77k] †<br>[22,38k,69k] † | **[13, 15, 16]** |
| | Coverage | 62.5%<br>62.5% | 100%<br>100% | **100%** |
| | BMC runtime (minutes) | N/A | N/A | **[19, 39, 71]** |

† If trace buffers are used for QED, then the trace lengths in terms of instructions are: for FFT with CFCSS-V [14, 92, 231], MMULT with CFCSS-V [14, 120, 192]. The trace lengths in terms of clock cycles are: for FFT with CFCSS-V [22, 112, 265], MMULT with CFCSS-V [22, 141, 222]. Other entries remain the same.

**Table 4**. Results comparing original tests (No QED) and QED family tests (CFTSS-V) for FFT (top values) and MMULT (bottom values), and Symbolic QED for CFTSS-V (initialized from a QED-consistent state obtained by running an FFT QED test). For bug traces, we report the length in instructions and clock cycles. We also report the BMC runtime in minutes.

| | | Original (No QED) | QED (CFTSS-V) | **Symbolic QED (CFTSS-V)** |
|---|---|---|---|---|
| **Control flow bug that results in a deadlock** | Bug trace length (instructions) | ~12B<br>~12B | 9<br>9 | **9** |
| | Bug trace length (clock cycles) | 12B<br>12B | 19<br>19 | **19** |
| | Coverage | 100%<br>100% | 100%<br>100% | **100%** |
| | BMC runtime (minutes) | N/A | N/A | **72** |

### *C. Effectiveness of Symbolic QED for CFCSS-V and CFTSS-V Transformations (for both Pre-silicon and Post-silicon validation)*

To evaluate Symbolic QED for CFCSS-V and CFTSS-V, we implemented bug scenarios from [14] (Table A1 of the Appendix) that affect the control flows of processor cores. Specifically, bug activation criteria 1-8 from Table A1.A and bug effect H from Table A1.B in the Appendix were implemented. In addition, we also implemented the bug from the case study presented in [14], i.e., a deadlock occurs when a specific sequence of 9 instructions (from the original test) are executed. The deadlock is implemented by stopping the pipeline of processor core so that the processor core does not fetch new instructions and does not commit instructions already in the pipeline. The bugs were implemented by modifying the RTL of the OpenSPARC T2. For Symbolic QED for CFTSS-V, we set C = 17 to get the property: "eventually, 17 instructions have to commit". As discussed in Sec. III.E, we empirically selected C = 17, the largest number of instructions the BMC tool could successfully analyze.

Tables 3 and 4 report the results of Symbolic QED for CFCSS-V and CFTSS-V, respectively. Since CFCSS-V QED targets bugs that result in incorrect control-flow and not livelocks / deadlocks, Table 3 only reports results corresponding to the 8 bug scenarios that result in incorrect control-flow (i.e., bug activation criteria 1-8 from Table A1.A and bug effect H from Table A1.B). Since CFTSS-V targets bugs that results in livelocks / deadlocks, Table 4 only reports results corresponding to the bug (described above) that results in a deadlock. Note that, the Symbolic QED results apply to pre-silicon verification as well (as in Table 1).

**Observation 4:** Symbolic QED correctly localizes control flow bugs in under 80 minutes and provides traces up to 6 orders of magnitude shorter than traditional post-silicon validation tests, and up to 5 orders of magnitude shorter than QED tests. These results include a bug that caused a deadlock.

Since both CFCSS-V and CFTSS-V target bugs that affect the control flows of processor cores, only the processor core was analyzed in the BMC tool. The Original (No QED) column shows results from running the original validation tests (FFT or MMULT) using end result checks to check the results of the test against pre-computed, known correct results. For the bug scenario that resulted in a deadlock (Table 4) we used a timeout of 10 seconds to detect if a deadlock occurred. Since the OpenSPARC T2 is designed to operate at 1.2 GHz, this corresponds to an error detection latency of 12 billion clock cycles.

Although none of the bug scenarios demonstrated explicitly simulate processor interrupt related bugs, the inclusion of CFCSS-V and CTCSS-V also enables Symbolic QED to detect and localize most interrupt related bugs. When an interrupt occurs, the processor branches to a pre-determined section of code to respond to the interrupt, and then branches back to continue executing from where it left off [36]. Note, therefore, that most processors executing an interrupt behave similarly to a processor executing any branch instruction followed by some code and then another branch back to the same location in the original code. As a result, a bug during the processor interrupt can have the following effects. The processor may execute the wrong code during the interrupt or it may return to the wrong place in the original code. Both of these cases are characterized by the branch to wrong address scenario that is already included in our bug examples. The results in Table 3 demonstrate that Symbolic QED for CFCSS-V correctly detects and finds traces for these types of bugs. Interrupt bugs can also cause the system to hang during the interrupt's execution in a deadlock or livelock which would be covered by the CFTSS-V implementation of Symbolic QED (Sec. III.E). The results in Table 4 again demonstrate the ability of this approach to detect and find instruction traces to localize these bugs.

### *D. Effectiveness of Symbolic QED for Bugs with Very Long Activation Sequences*

Some logic bugs can be difficult to activate, because they require a specific state that can only be reached by executing a long sequence of specific instructions (far beyond the BMC bound) or starting the BMC analysis from a specific rather than generic initial state. Such bugs may escape pre-silicon verification, and require extensive testing using QED tests for detection during post-silicon validation. Symbolic QED helps localize such difficult bugs (after detection during post-silicon validation using QED tests). The key idea is to use the "state" of the design at an "appropriate" point (e.g., prior to or immediately after bug detection by QED) as an initial state for Symbolic QED.

To illustrate our approach, we consider three bug scenarios in Table 5. Each activation criterion in Table 5.A requires long sequences. A previously described bug effect from [13, 14] is shown in Table 5.B. Using a generic (QED-consistent) initialization, Symbolic QED initially failed to find a trace (i.e., the BMC tool timed out) for all three bugs (e.g., during pre-silicon verification). A QED-transformed MMULT benchmark with *Inst_min* and *Inst_max* set to 50 detected all three bugs (during post-silicon validation). However, the QED test cannot localize these bugs. As discussed in Sec. III.F, QED can localize a bug to a processor core (if it is detected by a normal or store check). However, bugs detected by a load check can come from any other module in the entire design. For the bug examples shown in this section, the bug effect results in an outdated cache entry (due to a dropped invalidate signal) in the L2 cache.

Based on QED results obtained during post-silicon validation, two different strategies can now be used to initialize the BMC tool for Symbolic QED.

**Strategy 1.** This method uses QED-consistent values obtained from running QED tests to initialize flip-flops corresponding to architectural states (register values and memory contents). We make some (minor) modifications to QED tests for this purpose. First, we instrument the QED test with the CFTSS-V QED transformation from Sec. II.C (along with other QED transforms such as EDDI-V to quickly detect bugs).

**Table 5.A.** Activation criteria requiring very long activation sequences.

| | |
|---|---|
| Processor Core | 1. *R* registers must each contain a specific value *V*. |
| | 2. A specific sequence of *N* instructions must execute within *X* cycles. |
| | 3. A specific cache state. |

**Table 5.B.** Bug effect from [13, 14].

| | |
|---|---|
| Uncore Component | Next received cache coherence message dropped |

**Transformed Code**

```
          ...
SIGNATURE = BLOCK_X
<CFTSS-V Operation>
R1  = R2  + R3
R4  = R5  - R6
R7  = R1  - R4
R4  = R7  * R8
R17 = R18 + R19
R20 = R21 - R22
R23 = R17 - R20
R20 = R23 * R24
SIGNATURE = BLOCK_Y
<CFTSS-V Operation>
          ...
```

(a)

**CFTSS-V Operation**
For Thread *N*

```
CFTSS_V_SIG_COUNT_N++
IF (CFTSS_V_SIG_COUNT_N == STOP)
 {
   GOTO END
 }
```

(b)

**Figure 18.** CFTSS-V transformed code (a) inserted before each block of instructions with *Inst_min* = *Inst_max* = 4; and (b) pseudo-code for incrementing a count of CFTSS-V signatures and checking if the stopping point has been reached, on thread *N*.

Next, we enhance the tests to additionally *count* the total number of CFTSS signature updates, $S_E$ (Fig. 18), that have been performed by each thread. The number of signature updates provides a rough estimate of when the bug was detected during test execution. Finally, we also add a check that stops execution of a thread if its signature count reaches some value STOP. The reason for this check is explained later, but it is included in the test from the beginning so that the tests do not need to be modified later. Each thread has its own signature variable and corresponding signature count. CFTSS-V signatures are inserted at QED-consistent points in the test (where the architectural state is QED-consistent, Sec. III.C).

We run the QED test (using the modified CFTSS-V plus other error detecting transforms, e.g., EDDI-V) initially with STOP set to 0 (to ensure that the STOP condition is not triggered; the signature count associated with the first CFTSS-V check starts at 1). When the test detects a bug, the values of each thread's CFTSS-V signature counts are saved. Next, the post-silicon QED test is rerun, but this time with each thread's STOP value set to the CFTSS-V signature counts from the run that detected the bug. Note that, this run (to obtain the architectural states) does not require failure reproduction (i.e., the bug to be detected again). It is even possible that for some pathological cases, the STOP condition will not be reached on a thread (e.g., if the signature update count was only reached due to the thread's behavior in the presence of the bug). Our general assumption is that, near each thread's final signature update the design's architectural state is "close" to the one required to activate the bug.

After having rerun the test to each thread's STOP condition, the BMC tool is initialized with the final values in the registers and memory. Symbolic QED is then used to find a bug trace starting from this architectural state. It is possible that the bug activation occurred *before* this state (as shown in the following example) and that this final state once again requires too long of an instruction sequence for the BMC tool to find a bug trace. One can decrement the CFTSS-V signature counts of one or more threads in that case to initialize Symbolic QED to an earlier state in an attempt to find a trace from there. Multiple different initializations can be tried in parallel.

For example, consider a scenario using the first activation criterion in Table 5.A with $R$ = 30 and $V$ = -1, and the bug effect in Table 5.B. Since 30 instructions would be required to load the given value into 30 registers to activate this bug, Symbolic QED cannot find a trace with a generic initialization. During our 8-core QED transformed MMULT post-silicon test, an EDDI-V check detected an error, after which we stopped the QED test and read out the CFTSS-V signature counts for each core (one thread per core). The QED test detected an error after counting the following CTFSS-V signature updates on the cores: 1,501, 1,463, 1,471, 1,466, 1,475, 1,468, 1,472 and 1,470. To find the architectural states (register and memory values) at each saved signature count, we re-ran the QED test and saved the architectural state on each core after those signature counts.

For each partial instance, we initialized the BMC tool with the register and memory values from the saved architectural states and then ran Symbolic QED. The BMC tool timed out with this first attempt for each of the partial instances. We then proceeded to search through earlier architectural states by decrementing the signature update count of a thread and re-running the QED test to find that architectural state. Symbolic QED successfully found a bug trace with a partial instantiation of 2 cores, the crossbar and 2 cache banks when initialized with an architectural state corresponding to the following CFTSS-V signature update counts on the cores: 1,499 (instead of 1,501 earlier) on Core 1 and 1,463 on Core 2. This example provides an important insight. The MMULT QED test immediately loaded new register values after the bug was activated – by the time the error was detected, the architectural state of the design had changed significantly. Between bug activation and detection, the error(s) in the internal flip-flop(s) was (were) not yet observable at the architectural state level. Hence, with the initialization corresponding to CFTSS-V signature update counts obtained upon bug detection, Symbolic QED timed out.

However, when initialized with an architectural state corresponding to "slightly" earlier CFTSS-V signature update count(s), twenty-seven of the registers on Core 1 already contained -1, and Symbolic QED could activate the bug by deriving instructions that write -1 to the remaining three. Symbolic QED could also derive instructions that detected the error (well within the BMC bound).

With Strategy 1, it is possible that the BMC tool may time out because a bug requires too many instructions to be activated with only the architectural states initialized. In such cases, we use Strategy 2.

**Strategy 2.** Strategy 1 initializes BMC with architectural states only. It is possible that internal flip-flops and/or caches may have to be correctly initialized for BMC to find a solution within BMC bounds. In these cases, we rely on RTL simulation to generate logic values of internal states for BMC initialization. (High-level simulators such as SIMICS [38] or gem5 [39] may also be used for faster simulation speeds, but the internal states visible to such tools can be limited.) Given the complexity of existing designs, simulating the system from the start of the QED test is impractical. Instead, we initialize each thread at the start of the simulation to a QED-consistent architectural state. That architectural state for each thread can be obtained from the CFTSS-V signature counts, as in Strategy 1.

The purpose of RTL simulation is to resolve internal flip-flop values of the system (**ideally** leaving none as unknown or Xs). Each RTL simulation starts from the *power on reset* state, after which the registers and memory values of each thread are set to the desired architectural state from the selected CFTSS-V signature update. During simulation, each thread then executes the remaining code that followed the chosen signature update on that thread. Starting from the *power on reset* state reduces the occurrences of internal Xs (unknown states). We observed from 1,000 RTL simulations of our QED-transformed MMULT test, our QED-transformed FFT test, and a test composed of randomly generated instructions (each starting from a random architectural state on all 8 cores) that nearly all Xs were resolved in the OpenSPARC T2 SoC after 1,000 cycles. The Xs that remained were limited to invalid data array values such as in the L2 cache eviction control logic, where the write-back array contains Xs until a modified entry exists in the L2 cache to be written to memory. These Xs should not propagate within the design as their entries would not be accessed unless they were overwritten by a valid value.

This suggests that it should be sufficient to run the RTL simulation for at least 1,000 cycles on each thread, after the initial reset sequence. One way to ensure this is to start RTL simulation from an architectural state corresponding to 1,000 CFTSS-V signature updates before the final one. For our examples, this meant closer to 100,000 cycles of RTL simulation; the total simulation time was still under forty minutes (Table 6). We perform RTL simulation up to certain CFTSS-V signature count values for each thread (the specific values are determined similar to Strategy 1). At the end of RTL simulation, we save the full design state (e.g. a VCD file), and use this state to initialize the BMC tool. The BMC tool then attempts to find a bug trace (through partial instantiation). We assume that multi-threaded applications do not need to be "perfectly" synchronized to derive the initial state for BMC.

**Table 6:** Runtime (minutes) simulating the entire OpenSPARC T2 RTL for a given number of CFTSS-V signature updates on each of the 8 cores, starting each core from a QED-consistent register and memory initialization from a MMULT benchmark transformed with EDDI-V, PLC and CFTSS-V.

| **Signature Updates** | 100 | 200 | 500 | 1,000 |
|---|---|---|---|---|
| **Runtime (min.)** | 6 | 10 | 21 | 38 |

Multiple threads running in parallel are unlikely to reach QED-consistent states (and update CFTSS-V signatures) simultaneously. Thus, it is likely that the exact sequence of operations will be different during RTL simulation vs. post-silicon test (e.g., shared variables may be updated in different orders). Similar to Strategy 1, we assume this strategy brings us "close" to the design state that triggers the bug.

We now return to the discussion of our three examples. We already discussed the first example using activation criterion 1 from Table 5.A. Our second bug example uses activation criterion 2 from Table 5.A with $N$=25 and $X$=30 and the bug effect in Table 5.B. The activating sequence of 25 instructions is too long to be detected by a generic Symbolic QED initialization or Strategy 1 (as mentioned in Sec. III.E, our BMC tool could not analyze instruction sequences beyond 17 instructions). The 8-core QED-transformed MMULT test (with Inst_min and Inst_max set to 50) detected the error after counting the following CTFSS-V signature updates: 3,736, 3,684, 3,691, 3,693, 3,690, 3,688, 3,686 and 3,691. For Strategy 2, we reran the test and stopped at the following signature update counts: 2,736, 2,684, 2,691, 2,693, 2,690, 2,688, 2,686 and 2,691 (1,000 updates prior to bug detection). We captured the architectural state at this point to initialize RTL simulation. We then simulated the full RTL through 1,000 signature updates on each core to reach the final signature update counts above. We used the full design state at this point to initialize the BMC tool for Symbolic QED. Using a partial instance consisting of one processor core (Core 2), the crossbar and two cache banks, the BMC tool returned a one-instruction bug trace, loading an entry from the L2 cache to Core 2. Significantly, this trace could be repeated by initializing Cores 3 through 8 in the partial instance but could not be obtained using Core 1. Analyzing that instruction trace with Core 1 resulted in the correct entry being loaded from Core 1's L1 cache. This suggested that the corrupted entry loaded by the other Cores was correctly saved in Core 1's L1 cache, indicating a bug involving the L2 cache system. Searching through earlier initializations (by stopping RTL simulation at earlier signature update count(s)), this same trace propagating the error from the L2 cache could be found only when Core 1 was simulated to one of its last two signature update counts (i.e. either 3,735 or 3,736).

This (correctly) indicated that the bug was activated after signature update 3,734 (within the 100 instructions preceding signature update count 3,735, determined by Inst_min/Inst_max of the QED test). At this point, additional debugging effort would be required to narrow the activation sequence even further. One potential approach we plan to explore in future work would be to take small subsets of these 100 instructions and simulate both an original and a corresponding duplicate copy of them and use the resulting (QED-consistent) state to initialize Symbolic QED.

Our third bug example used activation criterion 3 from Table 5.A (also with the bug effect in Table 5.B) and required all of the L1 cache lines to be valid. It also could not be activated by the BMC tool using Strategy 1, as expected, since it requires a specific cache state. The 8-core MMULT QED test detected the bug after reaching the following signature update counts: 2,273, 2,241, 2,243, 2,236, 2,247, 2,239, 2,243 and 2,240. For Strategy 2, we loaded the RTL simulator with the architectural state corresponding to signature update counts: 1,273, 1,241, 1,243, 1,236, 1,247, 1,239, 1,243 and 1,240. We then ran RTL simulation up to the final signature update counts and initialized the BMC tool with that design state. Symbolic QED, using a partial instantiation of one core, the crossbar and two L2 cache banks returned a one-instruction trace, loading from the L2 cache to the core. Similar to bug example 2 (and as expected since this example used the same bug effect), the trace could only be found using Cores 2 to 8 but could not be found using Core 1. The bug trace was only produced when Core 1 reached one of its final four signature update counts, starting at 2,270. This (correctly) localized the bug's activation to the 100 instructions between signature updates 2,269 and 2,270 on Core 1.

**Observation 5:** The results in Table 7 show that, with a combination of QED post-silicon tests and Symbolic QED, it is possible to localize each of the very long activation sequence bugs with an activating instruction trace and a partial instantiation in under 60 minutes. Significantly, for these bugs, QED tests alone cannot provide this degree of localization.

**Table 7**. Results for bugs with long activation sequences: comparing original MMULT test (No QED), QED tests (EDDI-V, PLC), Symbolic QED, and QED tests (EDDI-V, PLC, CFTSS-V) followed by initialized Symbolic QED. For bug traces, we report the length in instructions and clock cycles for a bug using each activation criterion [1, 2, 3] in Table 5.A. We similarly report BMC runtimes for each in minutes. Bug activation criterion 1 used Strategy 1; criteria 2 and 3 used Strategy 2.

| | Original (No QED) | QED (EDDI-V, PLC) | Symbolic QED | **QED followed by Symbolic QED (initialized using Strategies 1 & 2)** |
|---|---|---|---|---|
| Bug trace length (instructions) | [2.1M] | [217, 170, 354] | [N/A] | **[5, 1*, 1*]** |
| Bug trace length (clock cycles) | [2.4M] | [260, 192, 426] | [N/A] | **[22, 5, 6]** |
| Coverage | 0% | 100% | 0% | **100%** |
| BMC runtime (minutes) | N/A | N/A | (Timeout) | **[55, 42, 47]** |
| Bugs Localized | 0% | 0% | 0% | **100%** |

*1-instruction Symbolic QED trace to load the error after the bug has been activated through initialization from RTL simulation (Strategy 2).

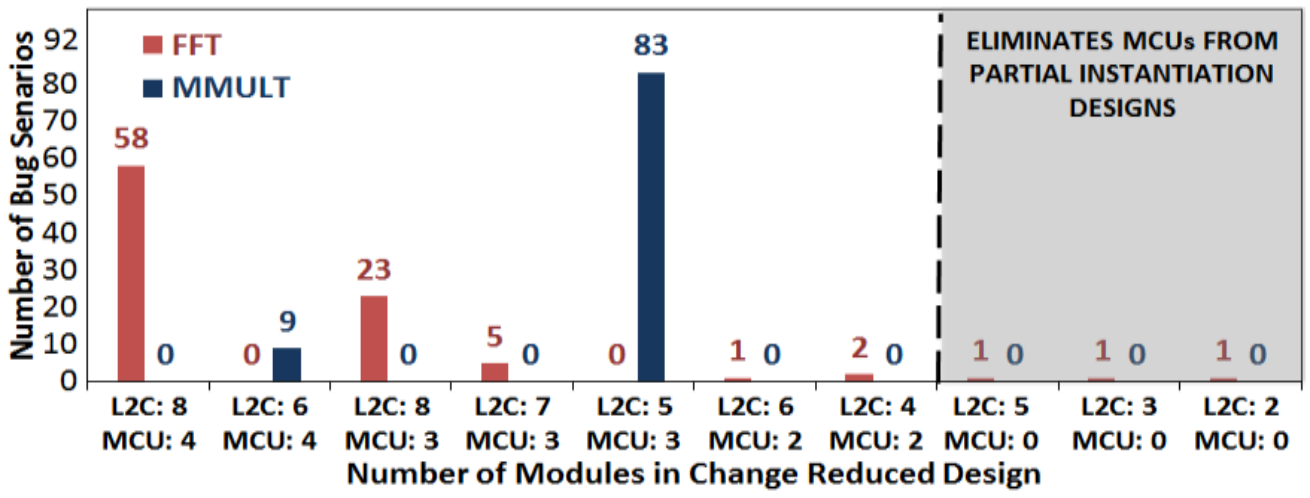

**Figure 19**. Change detector reduced design results for 92 bugs activated during the FFT and MMULT benchmark tests.

### E. *Usefulness of Change Detectors during Post-silicon Validation*

To gauge the usefulness of the change detectors in reducing the design size, we used the same 92 bug scenarios in OpenSPARC T2 and ran the QED transformed (EDDI-V and PLC) FFT benchmark along with a QED transformed (EDDI-V and PLC) matrix multiplication (MMULT) benchmark. With a change window of 1023 cycles, we observed that for these benchmarks, only the processor cores, crossbar, L2 caches and memory controllers were part of the reduced design. Other components such as the I/O modules and data management unit were removed from the design. For the 92 bugs, as shown in Fig. 19, the number of L2 caches (L2C) and memory controllers (MCU) also varied. However, only for 3 of the bugs (using the FFT benchmark) was the design size reduced enough to completely eliminate one set of these modules (the memory controllers) from the partial instantiation designs. We performed synthesis using Synopsys Design Compiler with the Synopsys EDK 32 nm library to calculate the area overhead of the change detectors on OpenSPARC T2 SoC.

**Observation 6:** We inserted change detectors on 1,067 signals with a total of 24,214 bits, thus requiring 24,214 change detectors for the entire design. This resulted in a 1.86% chip-level area overhead. However, given that the change detectors did not reduce the number of memory controllers or caches enough to eliminate a partial instantiation design for most of the bugs (Fig. 19), in this example we can remove the change detectors that only observe signals between these components. The number of signals monitored then drops to 899, removing several large data buses and requiring only 12,734 bits to be monitored. The area overhead now reduces to 0.98%. Thus the partial use of change detectors, generally on peripheral components that see intermittent activity, appears to be cost-effective; monitoring components such as caches and memory controllers that have high switching activity does not add significant value. The overhead is significantly less than the 4% overhead of reconfigurable logic for post-silicon debugging [7]. Furthermore, this approach avoids using trace buffers [11, 24, 38- 40].

## V. Related Work

The Symbolic QED technique in this paper mostly leverages prior work from QED [12, 13, 14], but there are important differences. Unlike Symbolic QED, QED alone does not directly localize bugs at a fine level of hardware granularity. As shown in Sec. IV, the bug traces obtained by QED can be very long (up to 5 orders of magnitude longer when no trace buffers are used) compared to Symbolic QED. Furthermore, Symbolic QED can be used during both pre-silicon and post-silicon, while QED is only an approach for a post-silicon validation. For bugs inside processor cores, Symbolic QED may be further enhanced by techniques such as self-consistency checking [43]. However, [43] addresses only processor core bugs. Our experience with bugs in commercial SoCs indicates that uncore components are an important source of difficult bugs [13, 14, 15].

The growing importance of post-silicon validation and debug has motivated recent publications on bug localization and bug trace generation. IFRA and the related BLoG [41, 42] techniques for post-silicon bug localization target processors only and the published results target electrical bugs. Their effectiveness for bugs inside uncore components is unclear. They also require manual efforts and additional hardware, unlike Symbolic QED.

Many post-silicon bug localization approaches rely on trace buffers and assertions. Sec. I discussed the inadequacy of these techniques (some of the heuristics for trace buffer insertion, e.g., restoration ratio and its derivatives, only work for logic bugs, since they use simulations to compute the logic values of signals that are not traced). In contrast, Symbolic QED doesn't require any trace buffers (or any additional hardware) or design-specific assertions and provides a very succinct and generic property to quickly detect and localize logic bugs.

BackSpace and its derivatives [24, 11, 12] provide a concrete bug trace once an error is detected or the system crashes by using formal methods to stitch together multiple short traces (or system states) into a longer trace. Some BackSpace derivatives require failure reproduction, which, as discussed in Sec. I and in [16, 22], is challenging due to Heisenbug effects [45]. nuTAB-BackSpace addresses some of the failure reproduction challenges but requires design-specific "rewrite rules" to determine if two similar states are equivalent. These rewrite rules have to be manually crafted by designers and require designer intuition, which may be difficult for large designs. Furthermore, the bug traces found may be very long, and unlike Symbolic QED, these techniques cannot reduce the length of the bug traces. Moreover, techniques that rely solely on formal methods for bug localization (e.g., [24, 11, 12, 25]) are not scalable to large designs (e.g. OpenSPARC T2). Some formal techniques require specific bug models (e.g., [25] which targets a specific model for electrical bugs) and may not work for logic bugs, since it is difficult to create models for all logic bugs [43].

Approaches that rely on detailed RTL simulations to obtain the internal states of a design are not scalable for large designs because full system RTL-level simulation of large designs is extremely slow, less than 10 clock cycles per second [23]. [46] presented a technique for post-silicon bug diagnosis, but it requires multiple detailed RTL simulations of the internal states of a design to guide the insertion of hardware structures for debugging. BuTraMin [38] is a pre-silicon technique for shortening the length of a bug trace. For use in post-silicon validation and debug of large designs, it requires massive simulations to capture logic values of all flip-flops in the system, which will be difficult. There may be opportunities to use such techniques *after* Symbolic QED localizes bugs and produces short bug traces (as demonstrated in this paper).

## VI. Conclusion

The Symbolic QED technique presented in this paper is a new structured and automated approach for logic bug detection and localization. It can be used to debug the design at any stage, both pre- and post-silicon. It detects logical bugs and provides a list of components that may contain the bugs along with the shortest instruction trace to activate the bug. Symbolic QED produces bug traces that are up to 6 orders of magnitude shorter than traditional post-silicon validation tests that rely on end-result-checks, and up to 5 orders of magnitude shorter than QED. It is completely automated, and does not require human intervention or additional hardware.

Symbolic QED is both effective and practical, as demonstrated on the OpenSPARC T2, where it correctly localized difficult logic bug scenarios that occurred during post-silicon validation of various commercial multicore SoCs. These difficult bug scenarios originally took many days or weeks of (mostly manual) debug work to localize.

Other formal techniques for debugging may take days or fail completely for large designs such as the OpenSPARC T2. As demonstrated in this paper, Symbolic QED is effective for bugs inside processor cores, bugs inside uncore components, as well as bugs related to power-management features and even bugs with long activation sequences. Symbolic QED is applicable to any SoC design if it contains at least one programmable processor core (a generally valid assumption for existing SoCs [3]).

There are several directions for future work. Symbolic QED can be expanded to: 1) demonstrate its ability to detect and localize bugs during pre-silicon or emulation-based verification; 2) localize electrical bugs during post-silicon validation (this paper's focus was on logic bugs); 3) perform full system-level bug localization; 4) perform diagnosis of manufacturing defects during system-level testing; 5) localize bugs in analog and mixed signal components; 6) investigate its applicability to software verification using an approach similar to [28]; 7) use a more general QED module that does not require all processor cores to start execution on the same cycle and starts duplication on a pseudo-instruction "QED" (instead of a control-flow instruction) and 8) use systematic techniques for initializing Symbolic QED during pre-silicon verification to further target bugs activated by long instruction sequences and difficult to reach design states (often only detected during post-silicon validation).

## References

[1] Adir, A., *et al.*, "Threadmill: A Post-Silicon Exerciser for Multi-Threaded Processors," *IEEE/ACM Design Automation Conf.*, 2011.

[2] Friedler, O., *et al.*, "Effective Post-Silicon Failure Localization Using Dynamic Program Slicing," *Proc. IEEE/ACM Design Automation Test in Europe*, pp. 1-6, 2014.

[3] Foster, H. D., "Trends in Functional Verification: A 2014 Industry Study," *Proc. IEEE/ACM Design Automation Conf.*, pp. 48-52, 2015.

[4] Keshava, J., N. Hakim, and C. Prudvi, "Post-silicon Validation Challenges: How EDA and Academia Can Help," *Proc. IEEE/ACM Design Automation Conf.*, pp. 3-7, 2010.

[5] Mitra, S., S. A. Seshia, and N. Nicolici, "Post-Silicon Validation Opportunities, Challenges and Recent Advances," *Proc. IEEE/ACM Design Automation Conf.*, pp. 12-17, 2010.

[6] Bohr, M., "The New Era of Scaling in an SoC World," *Proc. IEEE Solid-State Circuits Conf.*, pp. 23-28, 2009.

[7] Abramovici, M. "A Reconfigurable Design-for-Debug Infrastructure for SoCs," *Proc. IEEE/ACM Design Automation Conf.*, pp. 7-12, 2006.

[8] Nahir, A., *et al*., "Post-Silicon Validation of the IBM POWER8 Processor," *Proc. IEEE/ACM Design Automation Conf.*, pp. 1-6, 2014.

[9] Yerramilli, S., "Addressing Post-Silicon Validation Challenges: Leverage Validation & Test Synergy," Keynote, *IEEE Intl. Test Conf.*, 2006.

[10] Amyeen, M. E., S. Venkataraman and M. W. Mak, "Microprocessor System Failures Debug and Fault Isolation Methodology," *Proc. IEEE Intl. Test Conf.*, pp. 1-10, 2009.

[11] Reick, K., "Post-Silicon Debug – DAC Workshop on Post-Silicon Debug: Technologies, Methodologies, and Best-Practices," *IEEE/ACM Design Automation Conf.*, 2012.

[12] Hong, T., *et al.*, "QED: Quick Error Detection Tests for Effective Post-Silicon Validation," *Proc. IEEE Intl. Test Conf.*, pp. 1-10, 2010.

[13] Lin, D., *et al.*, "Quick Detection of Difficult Bugs for Effective Post-Silicon Validation," *Proc. IEEE/ACM Design Automation Conf.*, pp. 561-566, 2012.

[14] Lin, D., *et al.*, "Effective Post-Silicon Validation of System-on-Chips Using Quick Error Detection," *IEEE Trans. Computer Aided Design of Integrated Circuits Systems*, Vol. 33, No. 10, pp. 1573-1590, 2014.

[15] Lin, D., *et al.*, "Quick Error Detection Tests with Fast Runtimes for Effective Post-Silicon Validation and Debug," *IEEE Design Automation and Test in Europe Conf.*, 2015.

[16] De Paula, F. M., *et al.*, "TAB-BackSpace: Unlimited-Length Trace Buffers with Zero Additional On-Chip Overhead," *Proc. IEEE/ACM Design Automation Conf.*, pp. 411-416, 2011.

[17] Deutsch, S., and K. Chakrabarty, "Massive Signal Tracing Using On-Chip DRAM for In-System Silicon Debug," *Proc. IEEE Intl. Test Conf.*, pp. 1-10, 2014.

[18] El Mandouh, E., and A.G. Wassal, "Automatic Generation of Hardware Design Properties from Simulation Traces," *Proc IEEE Intl. Symp. Circuits and Systems*, pp. 2317-2320, 2012.

[19] Hangal S., *et al.*. "IODINE: A Tool to Automatically Infer Dynamic Invariants," *Proc. IEEE/ACM Design Automation Conf.*, pp. 775-778, 2005.

[20] Li, W., F. Alessandro, and S. A. Seshia, "Scalable Specification Mining for Verification and Diagnosis," *Proc. IEEE/ACM Design Automation Conf.*, 2010.

[21] Vasudevan, S., *et al.*, "GoldMine: Automatic Assertion Generation Using Data Mining and Static Analysis," *Proc. IEEE Design Automation and Test in Europe Conf.*, pp. 626-629, 2010.

[22] De Paula, F. M., A.J. Hu, and A. Nahir, "nuTAB-BackSpace: Rewriting to Normalize Non-Determinism in Post-Silicon Debug Traces," *Proc. Intl. Conf. on Computer Aided Verification*, pp. 513-531, 2012.

[23] Schelle, G., *et al.*, "Intel Nehalem Processor Core Made FPGA Synthesizable," *Proc. ACM/SIGDA Intl. Symp. Field Programmable Gate Arrays*, pp. 3-12, 2010.

[24] De Paula, F. M., *et al.*, "BackSpace: Formal Analysis for Post-Silicon Debug," *Proc. Formal Methods in CAD*, pp. 1-10, 2008.

[25] Zhu, C.S., G. Weissenbacher, and S. Malik, "Post-Silicon Fault Localisation Using Maximum Satisfiability and Backbones," *Proc. IEEE/ACM Formal Methods Computer-Aided Design*, pp. 63-66, 2011.

[26] Lin, D., *et al*., "A Structured Approach to Post-Silicon Validation and Debug Using Symbolic Quick Error Detection," in *2015 IEEE International Test Conference (ITC)*, 6-8 Oct. 2015

[27] Clarke, E., A. Biere, R. Raimi, Y. Zhu, "Bounded Model Checking Using Satisfiability Solving," *Formal* Methods *in System Design*, Vol. 19, No. 1, pp. 7-34, 2001.

[28] Campbell, K., D. Lin, S. Mitra, D. Chen, "Hybrid Quick Error Detection (H-QED): Accelerator Validation and Debug Using High-Level Synthesis Principles," *Proc. IEEE/ACM Design Automation Conf.*, pp. 1-6, 2015.

[29] D. J. Lu, "Watchdog Processors and Structural Integrity Checking," *IEEE. Trans. Comput.*, vol. 31, no. 7, pp. 681-685, Jul. 1982.

[30] Oh, N., *et al.*,"Control Flow Checking by Software Signatures," *IEEE Trans. on Reliability*, vol. 51, no. 1, pp. 111-122, Mar. 2002.

[31] J. P. Shen, and M. A. Schuette, "On-line Self-Monitoring Using Signatured Instruction Streams," *Proc. IEEE Intl. Test Conf.*, 1983, pp. 275–282.

[32] Sanchez, D., and C. Kozyrakis, "ZSIM: Fast and Accurate Microarchitectural Simulation of Thousand-Core Systems," *Proc. ACM Intl. Symp. Computer Architecture*, pp. 475-486, 2013.

[33] Shiu, P. H., Tan Y., Mooney, V. J. "A novel parallel deadlock detection algorithm and architecture." *Proceedings of the Ninth International Symposium on Hardware/Software Codesign (CODES 2001)*, IEEE, 2001.

[34] Available: http://www.opensparc.net

[35] Singh E., C. Barrett, S. Mitra, "Logic Bug Detection and Localization Using Symbolic Quick Error Detection," arXiv:[cs.OH], 2017

[36] Woo, S. C., *et al.*, "The SPLASH-2 Programs: Characterization and Methodological Considerations," *Proc. ACM/IEEE Intl. Symp. Computer Architecture*, pp. 24-36, 1995.

[37] Corbet, J., Rubini, A., Kroah-Hartman. G. "Interrupt Handler." *Linux Device Drivers. 3rd ed.* N.p.: O'Reilly Media, n.d. 258-87. Print.

[38] Chang, K., I. L. Markov, V. Bertacco, "Bug Trace Minimization," *Functional Design Errors in Digital Circuits*, Vol. 32, pp. 77-103, 2009

[39] P. S. Magnusson et al., "Simics: A full system simulation platform," in Computer, vol. 35, no. 2, pp. 50-58, Feb 2002.

[40] Binkert, N et. al. "The gem5 simulator", *SIGARCH Computer Architecture News* 39, 2 (August 2011), 1-7.

[41] Park, S.-B., T. Hong, and S. Mitra, "Post-silicon Bug Localization in Processors Using Instruction Footprint Recording and Analysis (IFRA)," *IEEE. Trans. Computer Aided Design Integrated Circuits System*, Vol. 28, No. 10, pp. 1545–1558, 2009.

[42] Park, S.-B., *et al.*, "BLoG: Post-Silicon Bug Localization in Processors Using Bug Localization Graph", *Proc. IEEE/ACM Design Automation Conf.*, pp. 368-373, 2010.

[43] Jones, R. B., C.-J. H. Seger, D. L. Dill, "Self-Consistency Checking," *Proc. Formal Methods in CAD*, pp. 159-171, 2005.

[44] Gray, J., "Why Do Computers Stop and What Can Be Done About It?" Tandem Computer, Tech. Report 85.7, PN 87614, 1985.

[45] http://www.itrs.net/Links/2009ITRS/Home2009.htm.

[46] DeOrio, A., D. S. Khudia, and V. Bertacco, "Post-Silicon Bug Diagnosis with Inconsistent Executions," *Proc IEEE Intl. Conf. Computer-Aided Design*, pp. 755-761, 2011.

## Appendix

A bug scenario is formed by pairing one bug activation criterion with one bug effect.

**Table A1.A.** Bug activation criteria from [13, 14].

| | |
|---|---|
| Uncore components | 1. Two stores within *X* clock cycles to different cache lines. |
| | 2. Two stores within *X* clock cycles to the same cache line. |
| | 3. Two stores within *X* clock cycles to adjacent cache lines. |
| | 4. Two cache misses within *X* cycles. |
| | 5. A sequence of loads and/or stores within *X* clock cycles. |
| Processor cores | 6. Data forwarding between pipeline stages. |
| | 7. Two branch instructions within *X* clock cycles. |
| Other | 8. A randomly chosen clock cycle. |

**Table A1.B.** Bug effects from [13, 14].

| | |
|---|---|
| Uncore components | A. Next received cache* coherence message dropped. |
| | B. Next received cache* coherence message delayed. |
| | C. Next store operation not allocated a cache* line. |
| | D. Next store update to cache* delayed by *Y* clock cycles. |
| | E. Next data accessed from cache* corrupted. |
| | F. Next data coming from main memory to cache* / core* corrupted. |
| | G. Processor core's* load value corrupted. |
| Processor cores | H. Core* jumps to incorrect (random) address in the next cycle. |
| | I. Error in decoding next instruction's operand inside core*. |
| | J. Processor core* incorrectly decodes next instruction to a NOP instruction. |

* Where activation criterion is satisfied.

**Table A2.A.** Power management bug activation criterion [15].

| ID | Description |
|---|---|
| 1 | When exiting from power-saving state. |

**Table A2.B.** Power management bug effects [15].

| Type | ID | Description |
|---|---|---|
| Uncore components | A | The value of the next load operation from data cache is corrupted to all 0's. |
| | B | Next load operation from data cache delayed (1 clock cycle) by cache controller. |
| | C | Data cache drops the next load operation. |
| | D | The value of the next load operation from main memory is corrupted to all 0's. |
| | E | Next load operation from main memory delayed (1 clock cycle) by memory controller. |
| | F | Next load request to main memory is dropped. |
| | G | Next load operation is delayed for 1 clock cycle by the interconnection network. |
| | H | Next load operation is corrupted to all 0's by the interconnection network. |
| | I | Next load operation is dropped by the interconnection network. |
| Processor cores | J | Processor jumps to a random address. |
| | K | Next instruction is corrupted to NOP |
| | L | The value of the next register read is corrupted to all 0's. |